\documentclass[superscriptaddress,showpacs,amssymb,10pt,reprint,aps,prd,longbibliography,nofootinbib,floatfix]{revtex4-2}

\usepackage{graphicx,epsfig,amssymb,times} 
\usepackage{amsmath,amsfonts}
\usepackage{bm}
\usepackage{epstopdf}
\usepackage{hyperref}
\usepackage[caption=false]{subfig}
\usepackage[usenames]{color}   
\usepackage[dvipsnames]{xcolor}
\usepackage[titletoc]{appendix}

\usepackage{soul}
\usepackage[normalem]{ulem}

\definecolor{darkblue}{rgb}{0,0,0.5}
\hypersetup{colorlinks=true, citecolor=darkblue, linkcolor=darkblue, urlcolor = darkblue}

\begin{document}

\title{Two shadows of a single black hole: \\ Vacuum birefringence phenomena within Einstein-nonlinear-electrodynamics}

	\author{Marco A. A. de Paula}
	\email{marcodepaula@ufpa.br}
	\affiliation{Programa de P\'os-Gradua\c{c}\~{a}o em F\'{\i}sica, Universidade Federal do Par\'a, 66075-110, Bel\'em, Par\'a, Brazil.}
	\affiliation{Faculdade de Ciências Naturais, Universidade Federal do Par\'a, Campus Universit\'ario do Tocantins-Cametá, 68400-000, Camet\'a, Par\'a, Brazil.}
	
	\author{Haroldo C. D. Lima}
	\email{haroldo.lima@ufma.br}
	\affiliation{Programa de P\'os-Gradua\c{c}\~{a}o em F\'{\i}sica, Universidade Federal do Maranh{\~a}o, Campus Universit{\'a}rio do Bacanga, 65080-805, S{\~a}o Lu{\'i}s, Maranh{\~a}o, Brazil.}
		
	\author{Pedro V. P. Cunha}
	\email{pvcunha@ua.pt}
	\affiliation{Departamento de Matem\'atica da Universidade de Aveiro and Centre for Research and Development  in Mathematics and Applications (CIDMA), Campus de Santiago, 3810-183 Aveiro, Portugal.}
	
	\author{Carlos A. R. Herdeiro}
	\email{herdeiro@ua.pt}
	\affiliation{Departamento de Matem\'atica da Universidade de Aveiro and Centre for Research and Development  in Mathematics and Applications (CIDMA), Campus de Santiago, 3810-183 Aveiro, Portugal.}
\affiliation{Programa de P\'os-Gradua\c{c}\~{a}o em F\'{\i}sica, Universidade Federal do Par\'a, 66075-110, Bel\'em, Par\'a, Brazil.}
	
	\author{Lu\'is C. B. Crispino}
	\email{crispino@ufpa.br}
	\affiliation{Programa de P\'os-Gradua\c{c}\~{a}o em F\'{\i}sica, Universidade Federal do Par\'a, 66075-110, Bel\'em, Par\'a, Brazil.}

\begin{abstract}

One of the main features of nonlinear electrodynamics is the existence of an effective geometry that describes the geodesic motion of photons. A detailed analysis of the properties of effective geometry is of utmost importance for a better understanding of nonlinear electrodynamics theories and their possible imprints on physics, especially in the context of black holes. We consider a nonlinear electrodynamics model that depends on the two electromagnetic scalar invariants and obtain that the motion of photons in nonlinear electrodynamics exhibits \textit{vacuum birefringence}, i.e., photons can propagate along two distinct paths, depending on their polarization. As a consequence of this phenomenon, we show that static black hole solutions sourced by nonlinear electrodynamics can admit two distinct unstable light rings, leading to the formation of two distinct shadows. Moreover, to explore the potential astrophysical relevance of our results, we also compare them with the astrophysical observations for the shadow radius of Sagittarius A*. We place upper limits on the charge-to-mass ratio of the nonlinear electrodynamics-sourced black hole. We also show that the motion of photons in this context can be interpreted as nongeodesic curves subjected to a four-force term from the perspective of an observer in the spacetime metric, generalizing previous results in the literature for nonlinear electrodynamics models that depend on a single electromagnetic scalar invariant.

\end{abstract}

\date{\today}

\maketitle

\section{Introduction}

Recent experimental developments in quantum field theory for very strong electromagnetic fields have provided some evidence for the light-by-light scattering~\cite{ATLAS2017,ATLAS2019,CMS2019} and the vacuum birefringence~\cite{RPM2016,LMC2017,A2021} (see also Refs.~\cite{RB2018,AF2023} for the state-of-the-art of these experiments). Both phenomena were first theoretically predicted in the mid-1930s~\cite{EK1935,HE1936} and are fundamental predictions of quantum field theory. These predictions show that a proper understanding of light requires the consideration of nonlinear effects for the electromagnetic field, in the appropriate regime. In astrophysics, e.g., vacuum birefringence is expected to be potentially detectable in the vicinity of some neutron stars, such as magnetars, due to their intense magnetic field. For instance, the magnetic field of SGR J1745-2900, a magnetar orbiting Sagittarius A$^{\star}$ (Sgr A$^{\star}$), is estimated to be around $10^{10}$ Tesla~\cite{JAK2013,magnetar}. 

The light-by-light scattering and vacuum birefringence phenomena are also closely related to models of nonlinear electrodynamics (NED), which can arise as effective nonlinear extensions of Maxwell electrodynamics. Since the 1930s, these models have been widely studied in the context of quantum electrodynamics, black hole (BH) physics, and string/M theories (see, e.g., Refs.~\cite{A2000,D2003,DPS2021,B2022} for further details). In the realm of BHs, one notable realization of NED models is that the motion of photons is affected by the nonlinearities of the electromagnetic field. The net effect of such nonlinearities for the propagation of photons can be obtained by computing the null geodesics of the so-called \textit{effective geometry}.

The concept of effective geometry in NED is commonly attributed to Jerzy Pleba\'nski, as introduced in Ref.~\cite{P1970}. He realized that, by considering the Born-Infeld (BI) electrodynamics~\cite{B1933,BI1934,B1934}, photons follow null geodesics in the background of an effective geometry, rather than the spacetime geometry~\cite{P1970}. He also found that the generalized BI model, which depends on the two electromagnetic scalar invariants, $F$ and $G$, exhibits no vacuum birefringence.\footnote{Throughout this paper, we call the NED model proposed by Born and Infeld that depends on $F$ and $G$~\cite{BI1934} the BI model, unless otherwise stated.} This result was also obtained independently by Guy Boillat in Ref.~\cite{B1970}, who investigated the dynamics of electromagnetic waves in the NED context. 

Gutiérrez \textit{et al.} generalized the results derived by Pleba\'nski to an arbitrary NED model~\cite{GDP1981}, and these results were also later rediscovered independently by Novello \textit{et al.}~\cite{MN2000}. In recent years, efforts have been made to clarify the meaning and implications of the effective geometry in NED-based regular BH (RBH) spacetimes (see, e.g., Ref.~\cite{B2023} for a review on RBHs and NED models). In particular, some investigations were carried out by performing a detailed geodesic analysis~\cite{JV2019,HR2020,AM2020,HJR2021,TAM2021,BL2021}, computing shadows~\cite{SS2019,SSO2019,AA2020,K2020,JR2020,ZH2021}, and  studying gravitational lensing effects~\cite{E2006,L2017,L20172,GAN2018,SS20192,WLF2015} of several electrically or magnetically charged NED-based RBH solutions. Moreover, the causality of NED-sourced spacetimes, particularly BHs, has also been addressed recently~\cite{SPL2016,TS2023,RW2024,MS2024,MAP2024d}.

Despite those advances in optical phenomena related to the effective geometry, there are still many open questions regarding its physical meaning. In Ref.~\cite{NBS2000}, the authors analyzed some consequences of the motion of photons in the Ayón-Beato-García (ABG) RBH spacetime~\cite{ABG1998}, which is the first electrically charged RBH solution based on NED. Among their main results, by analogy with electrostatic fields inside isotropic dielectrics, they showed that the effective geometry can be interpreted as if photons were governed by linear electrodynamics in the presence of a dielectric medium.

Recently, an alternative interpretation --- but equivalent --- for the effective geometry was provided in Ref.~\cite{MP2023}. The authors demonstrated that the null geodesics in the effective geometry are equivalent to nongeodesic curves subjected to a four-force term from the perspective of a timelike observer in the standard geometry. Therefore, according to an observer whose measurements are ruled by the spacetime metric, photons within NED are accelerated by a four-force related to the nonlinearities of the electromagnetic field. This interpretation is particularly useful when exploring the shadows and gravitational lensing of RBH geometries based on NED.

The results regarding the effective geometry presented in Refs.~\cite{NBS2000,MP2023} consider NED models that depend only on the electromagnetic scalar $F$, in the absence of external electromagnetic fields. Therefore, some optical phenomena, such as vacuum birefringence, are absent in this context. In this work, we find that, even in the presence of vacuum birefringence, the effective geometry can be interpreted as a nongeodesic curve subjected to a four-force term from the perspective of an observer in the spacetime metric. Through an analogy with the motion of charged particles in the spacetime metric, we also show that the four-force can be related to a four-current, which is associated with the nonlinearity of the electromagnetic field. 

Moreover, we consider the Euler-Heisenberg (EH) electrodynamics, proposed by Werner Heisenberg and Hans Euler in 1936~\cite{HE1936},\footnote{We emphasize that, although the Lagrangian resembles the EH form [cf.~\eqref{EH_Lagrangian}], here $\mu$ is not taken to be the quantum electrodynamics coefficient, but is treated as a generic free parameter. Consequently, the model does not describe the quantum electrodynamics coefficient, but serves as a generic NED suitable for exploring strong-field optical phenomena.} as a sample test to explore the relationship between gravitational lensing and vacuum birefringence. The EH electrodynamics provides an effective action that describes the nonlinear corrections to Maxwell's theory in vacuum, due to the effects of virtual electron-positron pairs. This description is related to the prediction of birefringence in vacuum and, for low-energy photons, to light-by-light scattering. Concerning our results, we show, in particular, that the vacuum birefringence introduces two distinct light rings associated with the formation of two different shadows. Moreover, the effective geometry related to the propagation of light rays typically increases the size of the shadow radius, when compared to the standard geometry, i.e., the geometry perceived by massless particles other than light in NED. We also compare our theoretical results with the observational data for the shadow radius of Sagittarius A$^{\star}$ (Sgr A$^{\star}$)~\cite{SV2023}, showing that extremely charged EH BHs are not compatible with the data. In addition, we also provide upper limits to the EH BH charge-to-mass ratio.

The remainder of this paper is organized as follows. In Sec.~\ref{sec:ef}, we present the field equations, the spacetime geometry, and the effective geometries. Our main results concerning the alternative interpretation of the effective metric are presented and discussed in Sec.~\ref{sec:eg4f}. In Sec.~\ref{sec:ehe}, we introduce the EH model. Our results for the shadows, gravitational lensing, and vacuum birefringence in the background of the magnetically charged EH BH are addressed in Secs.~\ref{sec:optical} and~\ref{sgl}. We end this paper with our final considerations in Sec.~\ref{sec:fr}. Throughout this paper we use natural units, for which $G = c = \hbar = 1$, and metric signature $(-,+,+,+)$.

\section{Background}\label{sec:ef}

The action that describes general relativity minimally coupled to  matter fields can be written as~\cite{HEL2006}
\begin{equation}
\label{generalaction}\mathrm{S} = \dfrac{1}{16\pi}S_{\rm{EH}}+S_{\rm{M}},
\end{equation}
where $S_{\rm{EH}}$ is the Einstein-Hilbert action and $S_{\rm{M}}$ is the action associated with the matter field. If $S_{\rm{M}}$ is related to a NED model, then we can rewrite Eq.~\eqref{generalaction} as
\begin{equation}
\label{S}\mathrm{S} = \dfrac{1}{16\pi} \int \sqrt{-g}\left[R-\mathcal{L}(F,G) \right] d^{4}x,
\end{equation}
in which $R$ is the Ricci scalar and $\mathcal{L}(F,G)$ is a gauge-invariant electromagnetic Lagrangian density, with $F$ and $G$ being the electromagnetic scalar invariants, defined by
\begin{equation}
\label{maxwellscalar}F \equiv F_{\mu\nu}F^{\mu\nu} \ \ \text{and} \ \ G \equiv F_{\mu\nu}\star F^{\mu\nu},
\end{equation}
respectively, where 
\begin{align}
\label{Fmunu}F_{\mu\nu} = \nabla_{\mu}A_{\nu} - \nabla_{\nu}A_{\mu},
\end{align}
is the Faraday tensor, $A_{\mu}$ is the four-vector potential, and $\star F^{\mu\nu}$ is the dual electromagnetic field tensor, namely
\begin{equation}
\label{dual}\star F^{\mu\nu} = \dfrac{1}{2}\epsilon^{\mu\nu\sigma\rho}F_{\sigma \rho}.
\end{equation}
The symbol $\star$ represents the Hodge dual operator and $\epsilon_{\mu\nu\sigma\rho}$ denotes the Levi-Civita tensor, which satisfies $\epsilon_{\mu\nu\sigma\rho}\epsilon^{\mu\nu\sigma\rho} = -4!$. The scalars $F$ and $G$ can also be written as
\begin{equation}
\label{scalareb}F = 2\left(B^{2}-E^{2}\right) \ \ \text{and} \ \ G = -4 \left(\vec{E}\cdot \vec{B} \right),
\end{equation}
where $\vec{E}$ and $\vec{B}$ are the electric and magnetic fields, respectively. For the nonlinear effects of the electromagnetic field to be appreciable, the magnitude of the fields must approach~\cite{G2008}
\begin{equation}
\label{threlecmag}E_{\rm{cri}} = 1.3\times10^{18}\, \dfrac{V}{m} \ \ \text{and} \ \ B_{\rm{cri}} = 4.4\times 10^{9} \, T,
\end{equation}
where $E_{\rm{cri}}$ and $B_{\rm{cri}}$ are the threshold values of the electric and magnetic fields, respectively.

By varying the action~\eqref{S} with respect to $A_{\nu}$, we obtain
\begin{equation}
\label{EFTC1}\nabla_{\mu}\left(\mathcal{L}_{F}F^{\mu\nu} + \mathcal{L}_{G}\star F^{\mu\nu}\right)  = 0,
\end{equation}  
where $\mathcal{L}_{F} = \partial\mathcal{L}/\partial F$ and $\mathcal{L}_{G} = \partial\mathcal{L}/\partial G$. Moreover, from the definition of $F_{\mu\nu}$ in Eq.~\eqref{Fmunu}, we get
\begin{equation}
\label{EFTC2}\nabla_{\mu}\star F^{\mu\nu} = 0.
\end{equation}
The Einstein-NED (ENED) field equations can be obtained by varying the action~\eqref{S} with respect to 
$g^{\mu\nu}$, which leads to
\begin{equation}
\label{E-NED_F}G_{\mu}^{\ \ \nu} = 8\pi T_{\mu}^{\ \ \nu} = 2\left[\mathcal{L}_{F}F_{\mu\alpha}F^{\nu\alpha}+\dfrac{1}{4}\left(G\mathcal{L}_{G}-\mathcal{L}\right)\delta_{\mu}^{\ \nu}\right].
\end{equation}
We notice that Eqs.~\eqref{S}-\eqref{E-NED_F} satisfy a correspondence with the standard linear electrodynamics in the weak field limit if 
\begin{equation}
\label{maxwelasym}\mathcal{L}(F) \rightarrow F, \ \ \mathcal{L}_{F} \rightarrow 1, \ \ \text{and} \ \  G = 0,
\end{equation}
for small $F$. In the far-field, the setup described above reduces to the standard Einstein-Maxwell theory in this limit~\cite{W1984}.

For concreteness, we consider a static and spherically symmetric spacetime, for which the line element is given by
\begin{equation}
\label{LE}ds^{2} = g_{\mu\nu}dx^{\mu}dx^{\nu} = -f(r)dt^{2}+f(r)^{-1}dr^{2}+r^{2}d\Omega^{2},
\end{equation}
where $d\Omega^{2} = d\theta^{2}+\sin^{2}\theta d\varphi^{2}$ is the line element of a 2-sphere of unit radius and $f(r)$ is the metric function, given by
\begin{equation}
\label{MF}f(r) = 1 - \dfrac{2\mathcal{M}(r)}{r}.
\end{equation}
The function $\mathcal{M}(r)$ is determined by the ENED field equations, given by Eq.~\eqref{E-NED_F}, and its integration in the whole space provides the Arnowitt-Deser-Misner mass $M$ of the BH. From now on, we refer to $g_{\mu\nu}$ as the \textit{spacetime metric}.

According to NED, the surfaces of discontinuity of the electromagnetic field propagate along an \textit{effective metric} $\bar{g}^{\mu\nu}$, which differs from the spacetime metric in general~\cite{P1970}. For the two-parameters Lagrangian density $\mathcal{L}(F,G)$, the effective metrics can be written as~\cite{GDP1981,MN2000}
\begin{align}
\nonumber \bar{g}^{\mu\nu}_{\pm} = & \mathcal{L}_{F}g^{\mu\nu}-4\big[\left(\mathcal{L}_{FF}+\Omega_{\pm}\mathcal{L}_{FG} \right)F^{\mu}_{\ \ \lambda}F^{\lambda\nu}+ \\
\label{eff_metric}& \left(\mathcal{L}_{FG}+\Omega_{\pm}\mathcal{L}_{GG}\right)F^{\mu}_{\ \ \lambda}\star F^{\lambda\nu}\big],
\end{align}
where
\begin{equation}
\label{Omega}\Omega_{\pm} = \dfrac{-\Omega_{2}\pm\sqrt{\left(\Omega_{2}\right)^{2}-4\Omega_{1}\Omega_{3}}}{2\Omega_{1}}.
\end{equation}
The functions $\Omega_{1}$, $\Omega_{2}$, and $\Omega_{3}$ are given by
\begin{align}
\Omega_{1} =\  & \mathcal{L}_{FG}\left(2F\mathcal{L}_{GG}-\mathcal{L}_{F}\right) + G\left(\mathcal{L}_{GG}^{2}-\mathcal{L}_{FG}^{2} \right), \\
\nonumber \Omega_{2} = \ & \left(\mathcal{L}_{F}+2G\mathcal{L}_{FG} \right)\left(\mathcal{L}_{GG}-\mathcal{L}_{FF} \right) + \\
& 2F\left(\mathcal{L}_{FF}\mathcal{L}_{GG}+\mathcal{L}_{FG}^{2} \right), \\
\Omega_{3} = \ & \mathcal{L}_{FG}\left(2F\mathcal{L}_{FF}+\mathcal{L}_{F}\right) + G\left(\mathcal{L}_{FG}^{2}-\mathcal{L}_{FF}^{2} \right),
\end{align}
respectively. Moreover, it can be shown that photons satisfy
\begin{equation}
\label{norm_NED}\bar{g}^{\mu\nu}_{\pm}k_{\mu}k_{\nu} = 0,
\end{equation}
where $k_{\mu}$ are null vectors with respect to the effective geometries $\bar{g}^{\mu\nu}_{\pm}$. By combining Eq.~\eqref{norm_NED} with the appropriate geodesic equations, it can be shown that the photon's path along the spacetime described by $\bar{g}^{\mu\nu}_{\pm}$ is indeed a geodesic curve (see, e.g., the Appendix A of Ref.~\cite{MN2000}). Therefore, in NED models, photons propagate along null geodesics of the effective geometries $\bar{g}^{\mu\nu}_{\pm}$, which are different from the geometry related to the spacetime metric $g^{\mu\nu}$. However, massless particles other than photons follow null geodesics of the spacetime metric given in Eq.~\eqref{LE}. In the remainder of this work, we use the term ``massless particles'' to refer to any particle that follows the null geodesics of the spacetime metric $g^{\mu\nu}$, while the term ``photon'' is used to refer to the perturbations of the electromagnetic field that are described by the null geodesics of the effective metric $\bar{g}^{\mu\nu}_{\pm}$.

The plus and minus signs in the effective metrics~\eqref{eff_metric} correspond to the two possible paths of light propagation, each one associated with a different polarization. In other words, when the Lagrangian, associated with a given NED model, depends on the two electromagnetic scalars $F$ and $G$, a vacuum birefringence phenomenon for light arises due to the nonlinearities of NED, except for the BI class. The existence of two different modes of propagation gives rise to intriguing phenomena in the shadow and gravitational lensing of RBHs sourced by NED. Additionally, we notice that NED models described by a Lagrangian that depends solely on $F$, $\mathcal{L}(F)$, do not exhibit such a vacuum birefringence phenomenon for light in the absence of external electromagnetic fields. In the remainder of this paper, we refer to the polarizations described by $\bar{g}^{\mu\nu}_{+}$ and $ \bar{g}^{\mu\nu}_{-}$ as polarizations $\mathcal{P}_+$ and $ \mathcal{P}_-$, respectively.

\section{Vacuum birefringence and four-force}\label{sec:eg4f}

In this section, we show that, from the perspective of an observer in the spacetime metric, the motion of photons can be interpreted as a nongeodesic curve subjected to a four-force term $\mathcal{F}^{\sigma}_{\ \ \mu\nu}$, generalizing the results presented in the Appendix B of Ref.~\cite{MP2023}. We also show that such a four-force can be associated with a four-current vector, which arises due to the nonlinearities of the electromagnetic field.

\subsection{Photons in NED as nongeodesic curves}

In order to derive an analytic expression for the four-force, we use the following identity, which is well-known in electrodynamics~\cite{NG2010},
\begin{equation}
\label{ident}F^{\nu}_{\ \ \alpha}\star F^{\alpha \mu} = -\dfrac{1}{4}Gg^{\mu\nu},
\end{equation}
and substituting in Eq.~\eqref{eff_metric}, we obtain
\begin{equation}
\label{eff_metric2}\bar{g}^{\mu\nu}_{\pm} = \mathcal{A}_{\pm}g^{\mu\nu}+\mathcal{B}_{\pm}h^{\mu\nu},
\end{equation}
where the quantities $\mathcal{A}_{\pm}$, $\mathcal{B}_{\pm}$, and $h^{\mu\nu}$ are given by
\begin{align}
\label{funcsABa}\mathcal{A}_{\pm} = \ & \mathcal{L}_{F}+\left(\mathcal{L}_{FG}+\Omega_{\pm}\mathcal{L}_{GG}\right)G, \\
\label{funcsABb}\mathcal{B}_{\pm} =\  & -4\left(\mathcal{L}_{FF}+\Omega_{\pm}\mathcal{L}_{FG}\right), \\
\label{funcsh}h^{\mu\nu} = \ & F^{\mu}_{ \ \ \rho}F^{\rho\nu},
\end{align}
respectively.

The covariant components of the effective geometry can be obtained from the identity $\bar{g}^{\mu\nu}_{\pm}\left( \bar{g}_{\nu\lambda}\right)_{\pm}=\delta^{\mu}_{\ \lambda}$. Thus, we get
\begin{align}
\label{EqApB2}\left(\bar{g}_{\mu\nu}\right)_{\pm}=m_{\pm} g_{\mu\nu}+n_{\pm}h_{\mu\nu},
\end{align}
where we defined 
\begin{align}
\label{aeq_1}m_{\pm} & \equiv -\dfrac{n_{\pm}\left(2\mathcal{A}_{\pm}-\mathcal{B}_{\pm}F \right)}{2\mathcal{B}_{\pm}}, \\
\label{aeq_2}n_{\pm} & \equiv -\dfrac{16\mathcal{B}_{\pm}}{8\mathcal{A}_{\pm}\left(2\mathcal{A}_{\pm}-\mathcal{B}_{\pm}F \right)-\mathcal{B}_{\pm}^{2}G^{2}}.
\end{align}

From the perspective of the effective geometries, the geodesic equation can be written as
\begin{align}
\label{EqApB3}\ddot{x}^\mu+\overline{\Gamma}^\mu_{\ \nu\sigma}\dot{x}^\nu\,\dot{x}^\sigma=0,
\end{align}
where $\overline{\Gamma}^\mu_{\ \nu\sigma}$ are the corresponding Christoffel symbols of the effective geometries, namely
\begin{equation}
\label{cs_eg} \left(\overline{\Gamma}^\mu_{\ \nu\sigma}\right)_{\pm} = \dfrac{1}{2}\bar{g}^{\mu\lambda}_{\pm}\left[\partial_{\nu}\left(\bar{g}_{\lambda\sigma}\right)_{\pm}+\partial_{\sigma}\left(\bar{g}_{\nu\lambda}\right)_{\pm}-\partial_{\lambda}\left(\bar{g}_{\nu\sigma}\right)_{\pm} \right].
\end{equation}
By inserting Eqs.~\eqref{eff_metric} and \eqref{EqApB2} into Eq.~\eqref{cs_eg}, and the resulting expression in Eq.~\eqref{EqApB3}, we obtain
\begin{align}
\label{sg_ff} \ddot{x}^\mu+\Gamma^\mu_{\ \nu\sigma}\dot{x}^\nu\,\dot{x}^\sigma =\left(\mathcal{F}^{\mu}_{\ \ \nu\sigma}\right)_{\pm}\dot{x}^{\nu}\dot{x}^{\sigma},
\end{align}
where
\begin{align}
\nonumber \left(\mathcal{F}^{\mu}_{\ \ \nu\sigma}\right)_{\pm} = & \ \dfrac{1}{16}\mathcal{B}_{\pm}n_{\pm}G^{2}\Gamma^\mu_{\ \nu\sigma} -\dfrac{1}{2}\big(\mathcal{A}_{\pm} g^{\mu\lambda}+ \\
\nonumber & \mathcal{B}_{\pm}h^{\mu \lambda} \big)\big[\partial_{\nu}\left(n_{\pm} h_{\lambda \sigma}\right)+\partial_{\sigma}\left(n_{\pm}h_{\nu\lambda}\right) - \\
\nonumber &   \partial_{\lambda}\left(n_{\pm}h_{\nu\sigma} \right)+g_{\lambda \sigma}\partial_{\nu}m_{\pm} +g_{\nu\lambda}\partial_{\sigma}m_{\pm}-\\
\label{four-force}&  g_{\nu\sigma}\partial_{\lambda}m_{\pm}\big]+\mathcal{B}_{\pm}m_{\pm}h^{\mu}_{\ \lambda}\Gamma^{\lambda}_{\ \nu \sigma},
\end{align}
and $\Gamma^\mu_{\ \nu\sigma}$ are the Christoffel symbols of the spacetime metric. We notice that Eq.~\eqref{sg_ff} represents a nongeodesic curve in the spacetime geometry, subjected to a four-force term appearing on the right-hand side. The analytic expression for this force is given in  Eq.~\eqref{four-force} and it acts on photons along their worldlines. Therefore, the motion of photons, even in the presence of vacuum birefringence, can be described by nongeodesic curves subjected to a four-force term, from the spacetime metric perspective. The interpretation that photons follow nongeodesic curves in the spacetime geometry complements the effective metric interpretation. These nonlinear interactions manifest as an effective force $\mathcal{F}^\mu_{\ \ \nu\sigma}$, as demonstrated in Eq.~\eqref{sg_ff}.

For a one-parameter Lagrangian density, i.e., $\mathcal{L} = \mathcal{L}(F,0)$, the first term in the right-hand side of Eq.~\eqref{four-force} is zero, and then the four-force reduces to
\begin{align}
\nonumber \mathcal{F}^{\mu}_{\ \ \nu\sigma} = \ & -\dfrac{1}{2}\left(\mathcal{L}_{F} g^{\mu\lambda}-4\mathcal{L}_{FF} h^{\mu \lambda} \right)\big[\partial_{\nu}\left(n h_{\lambda \sigma}\right)+ \\
\nonumber & \partial_{\sigma}\left(nh_{\nu\lambda}\right) - \partial_{\lambda}\left(nh_{\nu\sigma} \right)+g_{\lambda \sigma}\partial_{\nu}m+g_{\nu\lambda}\partial_{\sigma}m- \\
\label{4force}& g_{\nu\sigma}\partial_{\lambda}m\big]+n\Phi h^{\mu \lambda}\Gamma_{\lambda \nu \sigma},
\end{align}
where we drop the indexes $\pm$, $m = \mathcal{L}_{F}^{-1}$, and
\begin{align}
\label{n}n = & \ \dfrac{4\mathcal{L}_{FF}}{\mathcal{L}_{F}\Phi},\\
\label{Phi}\Phi = & \ \mathcal{L}_{F}+2F\mathcal{L}_{FF}.
\end{align}
Hence, we can interpret the first term in the right-hand side of Eq.~\eqref{four-force}, i.e., $\mathcal{B}_{\pm}n_{\pm}G^{2}\Gamma^\mu_{\ \nu\sigma}/16$, as a four-force term explicitly induced by the presence of the scalar invariant $G$ (or, if present, by the vacuum birefringence). In the Appendix~\ref{apxA}, we show that, by properly manipulating Eqs.~\eqref{4force}-\eqref{Phi}, we obtain the results derived in Ref.~\cite{MP2023} for static and spherically symmetric electrically charged RBHs.

\subsection{On the source of the four-force term}

For completeness, we also suggest a possible interpretation for the existence of the four-force in terms of a four-current vector associated with the motion of photons. We begin with the case $\mathcal{L} = \mathcal{L}(F,0)$, and notice that we can rewrite Eq.~\eqref{EFTC1} as
\begin{equation}
\label{NEDE_2}\nabla_{\mu}F^{\mu\nu} = 4\pi j^{\nu}_{\rm{NED}}(F,0),
\end{equation}
where $j^{\nu}_{\rm{NED}}(F,0)$ is given by
\begin{equation}
\label{eff_current}j^{\nu}_{\rm{NED}}(F,0) \equiv -\dfrac{\mathcal{L}_{FF}}{2\pi \mathcal{L}_{F}}F^{\mu\nu}F^{\alpha\beta}\nabla_{\mu}F_{\alpha\beta},
\end{equation}
and can be interpreted as an effective four-current density related to the nonlinearities of the electromagnetic field. Therefore, as the gravitational field in general relativity, the electromagnetic field self-interacts within the NED framework. Notice that if the limits given by Eq.~\eqref{maxwelasym} hold, then $j^{\nu}_{\rm{NED}}(F,0)$ vanishes, and we obtain the Maxwell results, i.e., $\nabla_{\mu}F^{\mu\nu} = 0$.

To investigate if $j^{\nu}_{\rm{NED}}(F,0)$ is conserved in curved spacetimes, we take the covariant derivative of Eq.~\eqref{NEDE_2}, leading to
\begin{equation}
\label{cons2}\nabla_{\nu}\nabla_{\mu}F^{\mu\nu} = 4\pi \nabla_{\nu}j^{\nu}_{\rm{NED}}(F,0).
\end{equation}
Since $F^{\mu\nu}$ is an antisymmetric tensor, we have
\begin{equation}
\label{nabla}\nabla_{\nu}\nabla_{\mu}F^{\mu\nu} = \dfrac{1}{2}\left[\nabla_{\nu},\nabla_{\mu}\right]F^{\mu\nu},
\end{equation}
which can be written as~\cite{IV2022}
\begin{equation}
\label{nabla2}\nabla_{\nu}\nabla_{\mu}F^{\mu\nu} = -R_{\mu\nu}F^{\mu\nu}+R_{\nu\mu}F^{\mu\nu},
\end{equation}
where $R_{\mu\nu}$ is the Ricci tensor. The Ricci tensor is symmetric, consequently $R_{\mu\nu} = R_{\nu\mu}$, implying that $\nabla_{\nu}\nabla_{\mu}F^{\mu\nu} = 0$. Therefore, the left-hand side of Eq.~\eqref{cons2} is zero, resulting in
\begin{equation}
\label{cons}\nabla_{\nu}\,j^{\nu}_{\rm{NED}}(F,0) = 0,
\end{equation}
and we see that $j^{\nu}_{\rm{NED}}(F,0)$ is conserved in curved spacetimes.

For the case $\mathcal{L} = \mathcal{L}(F,G)$, we obtain
\begin{align}
\label{generalcase}\nabla_{\mu}F^{\mu\nu} =   4\pi j^{\nu}_{\rm{NED}},
\end{align}
where $j^{\nu}_{\rm{NED}} \equiv j^{\nu}_{\rm{NED}}(F,G)$ is given by
\begin{align}
\nonumber j^{\nu}_{\rm{NED}} \equiv \ & -\dfrac{1}{2\pi \mathcal{L}_{F}}\bigg[\mathcal{L}_{FF}F^{\mu\nu}F^{\alpha\beta}\nabla_{\mu}F_{\alpha\beta} + \mathcal{L}_{GG}\times \\
\label{eff_current2} & \star F^{\mu\nu}\star F^{\alpha\beta}\nabla_{\mu}F_{\alpha \beta} -\dfrac{1}{2}\mathcal{L}_{G}\nabla_{\mu}\left(\star F^{\mu\nu} \right)\bigg],
\end{align}
and we can show that $j^{\nu}_{\rm{NED}}$ is also conserved by following the same steps given by Eqs.~\eqref{nabla}-\eqref{cons}. Moreover, notice also that the first term in the right-hand side of Eq.~\eqref{eff_current2} is $j^{\nu}_{\rm{NED}}(F,0)$.

\section{Sample model: EH electrodynamics}\label{sec:ehe}
\label{EH_Model_Sec}
In this section, we aim to investigate a concrete example of a BH spacetime sourced by NED, focusing on the relation between vacuum birefringence, shadows, and gravitational lensing. Specifically, we consider the Euler-Heisenberg (EH) model, described by the following Lagrangian~\cite{NY2022}:
\begin{equation}
\label{EH_Lagrangian}\mathcal{L}(F,G) = F-\mu \left(F^{2}+\dfrac{7}{4}G^{2}\right),
\end{equation}
where $\mu$ is considered as a generic free parameter. However, the Lagrangian~\eqref{EH_Lagrangian} also describes the low energy limit of quantum electrodynamics when $\mu$ is given by\footnote{Throughout this paper, we have adopted the natural units. Therefore, to obtain the correct value of $\mu$, we need to restore the units $c$ and $\hbar$. Following Ref.~\cite{JS1951}, one can show that $\mu$ with the units restored is given by
\begin{equation}
\label{muvalue}\mu_{\text{QED}} = \dfrac{2\alpha^{2}}{45}\left(\dfrac{\hbar^{3}}{4\pi\,m_{e}^{4}c^{5}} \right),
\end{equation}
in Gaussian unit and it has the dimension of meter$^{3}$/joule.}
\begin{equation}
\label{finestructure}\mu_{\text{QED}} = \dfrac{2\alpha^{2}}{45}\frac{1}{4\pi{m_{e}^{4}}} \approx 0.074 \times (M_{10k\odot})^2,
\end{equation}
with $\alpha$ and $m_{e}$ being the fine-structure constant and the electron mass, respectively. For convenience, we express $\mu_{\text{QED}}$ normalized by the mass of a $10^4\,M_\odot$ BH, denoted by $M_{10k\odot}$. Remarkably, even when $\mu$ is considered to be the quantum electrodynamics parameter, the NED contributions are relevant for BHs with masses smaller than $M_{10k\odot}$, as $\mu_{\text{QED}}$ gives a small but non-negligible contribution. Moreover, notice that for $\mu = 0$, we recover linear electrodynamics, i.e., $\mathcal{L}(F,G) = F$.

The setup is given by the ENED field equations~\eqref{E-NED_F}, with the line element~\eqref{LE}, and metric function~\eqref{MF}, for a purely magnetically charged NED source. We choose to work with magnetically charged Euler-Heisenberg black hole (EH BH), but electrically charged ones are also suitable for our applications. Details and derivation of the electric counterpart can be found in Ref.~\cite{BL2021}. 

Hence, the only non-null components of the electromagnetic field tensor are given by $F_{23}$ and $F_{32}$, which are related by $F_{23} = -F_{32} = Q\sin\theta$. From the $G_{0}^{\ 0}$- and $G_{2}^{\ 2}$-components of the ENED field equations~\eqref{E-NED_F}, we find
\begin{align}
\label{G00}&\dfrac{\mathcal{M}^{\prime}(r)}{r^{2}}  = \dfrac{1}{4}\mathcal{L}(F,G), \\
\label{G22}&\dfrac{\mathcal{M}^{\prime\prime}(r)}{r}= \dfrac{1}{2}\mathcal{L}(F,G)-F\mathcal{L}_{F}, 
\end{align}
respectively, where the prime symbol denotes differentiation with respect to the radial coordinate $r$. By solving Eq.~\eqref{G00} for $\mathcal{M}(r)$, we get
\begin{align}
\label{Q}\mathcal{M}(r) = C - \dfrac{Q^{2}}{2r} + \dfrac{\mu Q^{4}}{5r^{5}},
\end{align}
where $C$ is an integration constant. Since $\lim_{r \rightarrow \infty} \mathcal{M}(r) = M$, we get $C = M$. Thereafter, the EH metric function yields
\begin{equation}
\label{MF_EH}f(r) = 1 - \dfrac{2M}{r} + \dfrac{Q^{2}}{r^{2}}\left(1-\dfrac{2\mu}{5}\dfrac{Q^{2}}{r^{4}} \right).
\end{equation}
We obtain the Reissner-Nordström (RN) spacetime when $\mu = 0$, and the Schwarzschild geometry is recovered for $Q = 0$.

The horizons can be found from $f(r) = 0$, but the roots lead to nonelucidating equations and we decided to not exhibit them here. The extreme charge case can be obtained by solving $f(r)$ and $f^{\prime}(r) = 0$ simultaneously, yielding
\begin{widetext}
\begin{align}
\label{rext}r_{\rm{ext}} & = \dfrac{1}{6}\left(5M\pm\sqrt{25M^{2}-24Q^{2}}\right), \\
\label{muext}\mu_{\rm{ext}} & = \dfrac{5\left[-25 M \left(125 M^4-225 M^2 Q^2+108 Q^4\right)+216Q^{6} +M\left(6r_{\rm{ext}}-5M \right) \left(625 M^4-825 M^2 Q^2+216Q^4\right)\right]}{2916 Q^4},
\end{align}
\end{widetext}
which are the extreme values of the radial coordinate $r$ and the EH parameter $\mu$, namely, $r_{\rm{ext}}$ and $\mu_{\rm{ext}}$, respectively. We notice that, in the derivation of the EH BH solution, $\mu$ is interpreted as the free parameter of the model, which is fixed in the Lagrangian, and not as a quantity associated with a Gauss law, as the electric charge $Q$. Therefore, here we derive a family of EH BH solutions characterized by the parameters $M$, $Q$, and $\mu$, with the latter being a free parameter of the model instead of the constant given by Eq.~\eqref{muvalue}. Furthermore, obtaining $Q$ as a function of $\mu$ and $M$ leads to nonelucidating expressions. Thus, we opted to show $\mu$ as a function of $Q$ and $M$.

\begin{figure}[!htbp]
\begin{centering}
    \includegraphics[width=\columnwidth]{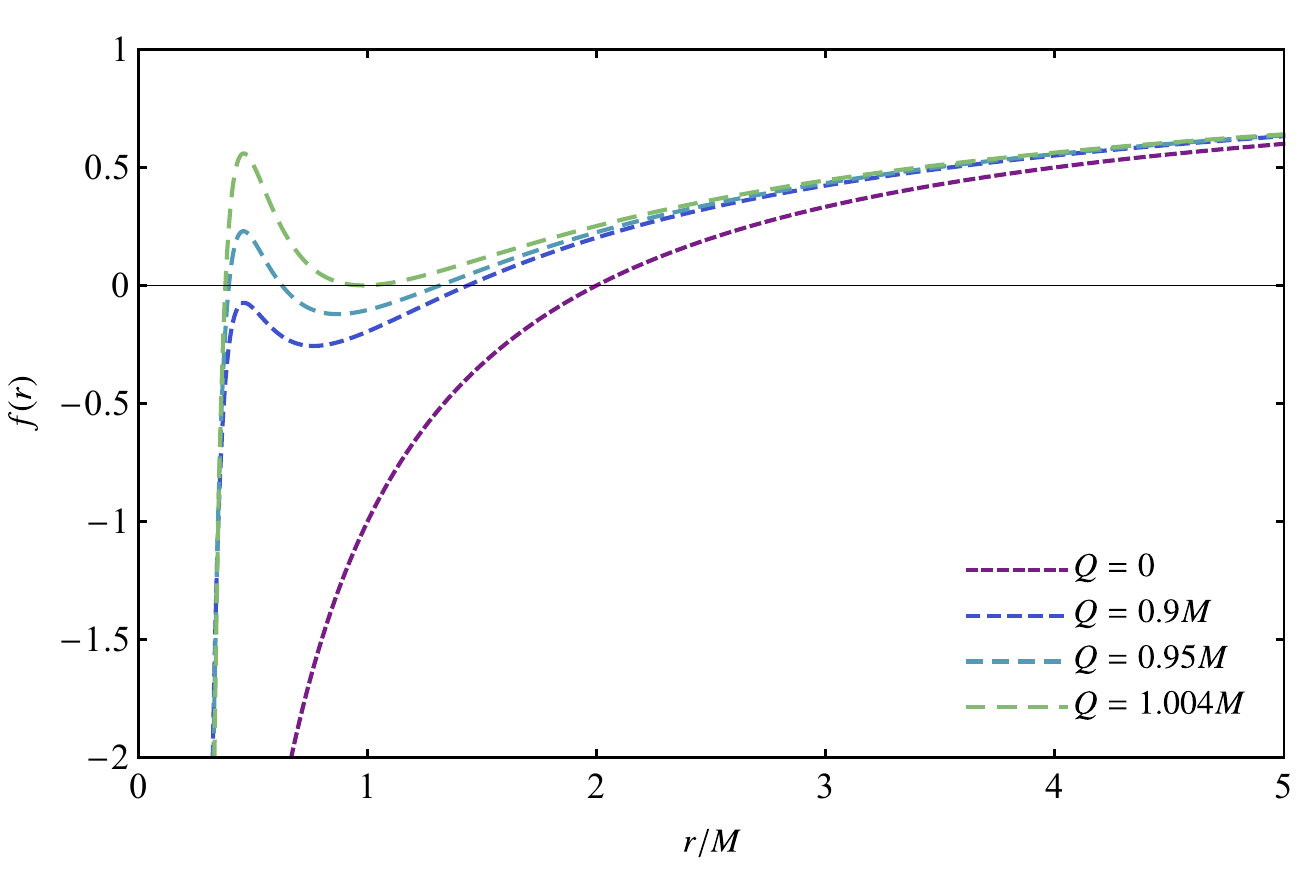}
    \caption{Metric function of the EH BH solution, considering $\mu = 0.02M^{2}$ and distinct values of $Q/M$, as a function of $r/M$. For this case, $r_{\rm{ext}} = 0.98222 M$ and $Q_{\rm{ext}} = 1.0042M$. We also exhibit the Schwarzschild case, $Q = 0$, for comparison purposes.}
    \label{metricfunction}
\end{centering}
\end{figure}

In this context, BH solutions exist for 
\begin{equation}
Q^{2} \leq \dfrac{25M^{2}}{24} \ \ \text{and} \ \ \mu \leq \dfrac{25M^{2}}{324},
\end{equation}
for which we may have up to three horizons, namely one event horizon $r_{h}$ and two inner horizons. At the extreme value of $Q/M$, the event horizon (largest positive root) is given by $r_{\rm{ext}} = 5M/6$. To illustrate this behavior, in Fig.~\ref{metricfunction} we display the metric function of the EH BH solution for $\mu = 0.02M^{2}$ and distinct choices of $Q/M$. We see that for $Q = Q_{\rm{ext}}$, we have an extreme EH BH with one event horizon and one inner horizon. For $Q < Q_{\rm{ext}}$, we have EH BHs with up to three horizons, depending on the ratio $Q/M$. Moreover, the spacetime is singular at the core and the Schwarzschild BH is obtained when $Q = 0$. We recommend Refs.~\cite{BL2021,AA2020,YT2001} for more details on the main properties of the EH BH spacetime.

We point out that in the NED framework it is relatively common to obtain BH solutions with BH charge-to-mass ratios satisfying $Q/M > 1$ while still admitting a well-defined event horizon. This is in contrast to the RN solution, where $Q/M > 1$ necessarily leads to a naked singularity. To name a few examples of NED-sourced BH spacetimes with this distinct feature, we mention one of the Ayón-Beato-García RBH geometries~\cite{ABG1999-2}, the Irina Dymnikova solution~\cite{D2004}, some of the RBH solutions derived in Ref.~\cite{BV2014}, and the RBH spacetime presented in Ref.~\cite{M2015}. We also note that in scenarios where $Q > Q_{\rm{ext}}$, the EH BH appears to have an event horizon, in contrast to what we observe in the RN spacetime. Here we restricted our analysis to BH solutions satisfying $Q \leq Q_{\rm{ext}}$, following the standard practice in the literature. We intend to thoroughly investigate the properties of the EH solution in scenarios where $Q > Q_{\rm{ext}}$ elsewhere.

The effective metrics can be obtained from Eq.~\eqref{eff_metric}. Since for EH Lagrangian density $\mathcal{L}_{FG} = 0$, the effective metrics can be put in a simple form, namely
\begin{align}
\label{gamma1}\bar{g}^{\mu\nu}_{+} & = \left(1+5\mu F\right)g^{\mu\nu}+14\mu F^{\mu}_{\ \lambda}F^{\lambda\nu}, \\
\label{gamma2}\bar{g}^{\mu\nu}_{-} & = \left(1-2\mu F\right)g^{\mu\nu}+8\mu F^{\mu}_{\ \lambda}F^{\lambda\nu},
\end{align}
where we neglected terms of $\mathcal{O}\left[\mu^{2}\right] $ in the effective metrics to be consistent with the EH NED source (proportional to $\mu$). 

The line elements of the effective metrics, for each polarization, can be written as
\begin{align}
\nonumber d\bar{s}^{2}_{\pm} & \equiv \left(\bar{g}_{\mu\nu}\right)_{\pm}dx^{\mu}dx^{\nu}, \\
\label{LE_EF}& =\dfrac{1}{G_{1}^{\pm}(r)}\left[-f(r)dt^{2}+\dfrac{1}{f(r)}dr^{2}\right]+\dfrac{r^{2}}{G_{2}^{\pm}(r)}d\Omega^{2},
\end{align}
in which we defined
\begin{align}
&G_{1}^{+}(r)  \equiv 1+\dfrac{10\mu Q^{2}}{r^{4}}, \\
&G_{2}^{+}(r) = G_{1}^{-}(r)  \equiv 1-\dfrac{4\mu Q^{2}}{r^{4}}, \\
&¨G_{2}^{-}(r)  \equiv 1-\dfrac{12\mu Q^{2}}{r^{4}}.
\end{align}
The functions $G^{\pm}_{j}(r)$, with $j = (1,2)$, are the so-called magnetic factors, which reduce to unity for $\mu = 0$ and/or $Q = 0$.

\begin{figure}[!htbp]
\begin{centering}
    \includegraphics[width=\columnwidth]{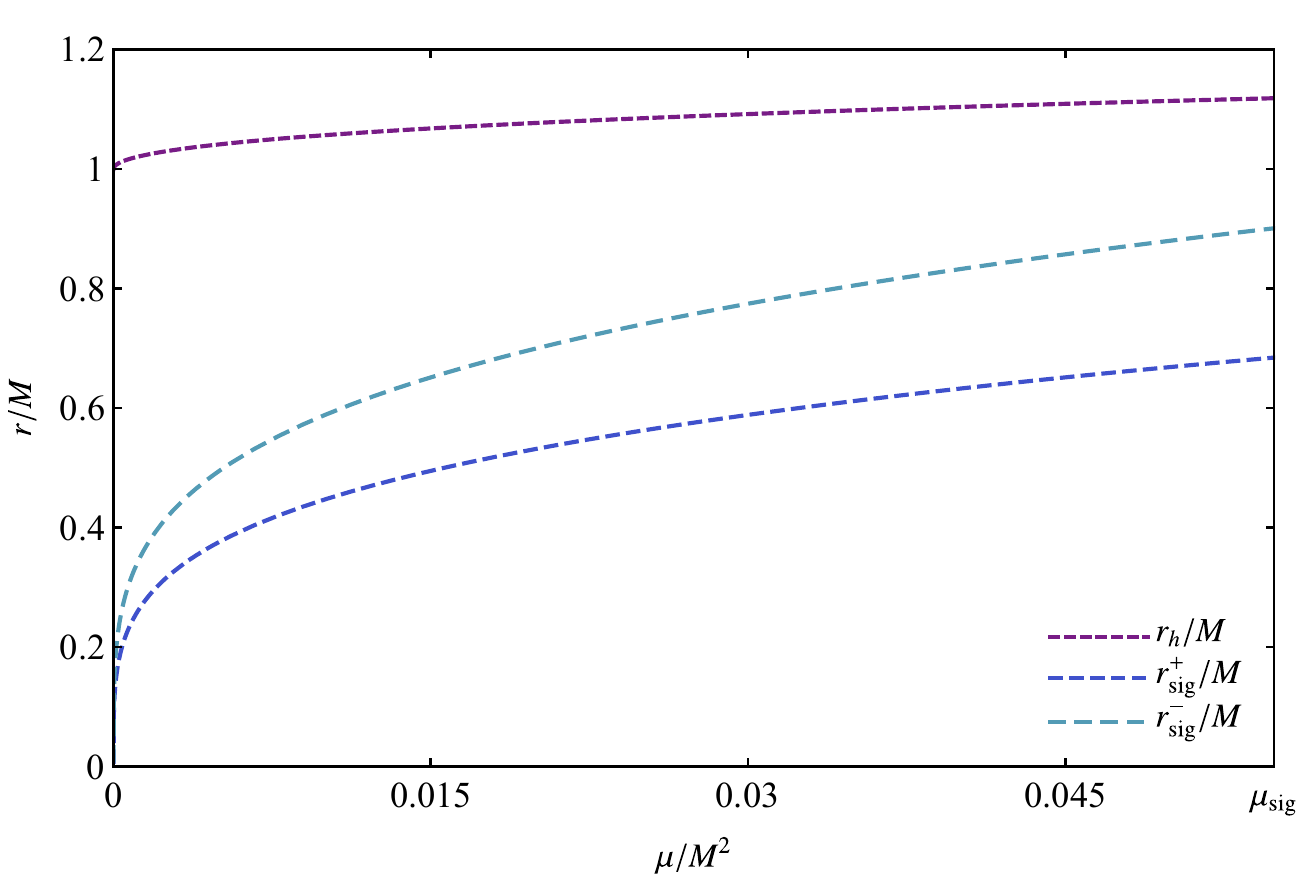}
    \caption{Comparison between the location of the event horizon with that of the signature radii, considering $Q = M$, as functions of $\mu/M^{2}$. For this case, $\mu_{\rm{ext}} = 40M^{2}/729$ and $Q_{\rm{ext}} = 1.0129M$.}
    \label{effradius}
\end{centering}
\end{figure}

As in other NED-based BHs~\cite{AA2020,MP2023}, it is possible that some components of the metric tensor change sign, which would represent a change in the signature of the effective metrics. To ensure that the effective metrics do not flip their signature, we define an effective radius for each polarization of the electromagnetic field, dubbed as signature radius, which are given by the zeros of $G^{-}_{j}(r) = 0$, namely
\begin{equation}
\label{effrad}r_{\text{sig}}^{+} \equiv \sqrt[4]{4\mu Q^{2}} \ \ \text{and} \ \ r_{\text{sig}}^{-} \equiv \sqrt[4]{12\mu Q^{2}},
\end{equation}
respectively (recall that $G_{2}^{+}(r) = G_{1}^{-}(r)$, hence the zeros of these magnetic factors coincide). For $r < r_{\rm{sig}}$, the signature of the effective metrics changes. In Fig.~\ref{effradius}, we exhibit a typical situations where the signature radii satisfy $r_{\rm{sig}} < r_{h}$. In general scenarios, this is not true, but throughout this work we always choose $Q$ and $\mu$ in a such way that $r_{h} > r_{\rm{eff}}$, since we are interested in optical effects that occur outside the event horizon. We also point out that the corresponding Kretschmann scalar of the EH BH geometry, computed in the standard geometry, is regular across the signature radius.

\section{Light rings and the effective metrics}\label{sec:optical}

In this section, we present the equations of motion for photons, according to each polarization of the electromagnetic field, discussing the existence of light rings (LRs) in this context. In view of the spherical symmetry, we consider the motion in the equatorial plane, without loss of generality.

\subsection{LRs and critical impact parameter}
\label{Subsec:VA}

To analyze the geodesic motion, according to each effective geometry, we apply the Hamiltonian formalism in curved spacetimes. The Hamiltonian that provides the equations of motion for photons in the effective metrics is given by
\begin{align}
\nonumber \mathrm{H}_{\pm} & \equiv \dfrac{1}{2} \bar{g}^{\mu\nu}_{\pm}\overline{p}_{\mu}\bar{p}_{\nu},\\
\label{H_EG}& = \dfrac{1}{2}\left[G^{\pm}_{1}(r)\left(-\dfrac{\bar{p}_{t}^{2}}{f(r)} + f(r)\,\bar{p}_{r}^{2}\right) + \dfrac{G^{\pm}_{2}(r)\bar{p}_{\varphi}^{2}}{r^{2}}\right],
\end{align}
where $\bar{p}_\mu$ are the components of the four-momentum of photons (we use $p_{\mu}$ for massless particles in the standard geometry). Since $\bar{p}_{\mu} = \bar{g}_{\mu\nu}\dot{x}^{\nu}$, we find
\begin{align}
\label{eqm1_EG}\dot{t} & = \dfrac{G^{\pm}_{1}(r)E}{f(r)}  ,\\
\label{eqm2_EG}\dot{r} & = f(r)G^{\pm}_{1}(r) \bar{p}_{r}  ,\\
\label{eqm3_EG}\dot{\varphi} & =  \dfrac{G^{\pm}_{2}(r) L}{r^{2}}.
\end{align}
Notice that we defined the constants of motion $\bar{p}_{t} \equiv -E$ and $\bar{p}_{\varphi} \equiv L$, where $E$ and $L$ are the energy and angular momentum of photons, respectively. These constants of motion arise due to the symmetries of the Hamiltonian~\eqref{H_EG} on the coordinates $t$ and $\varphi$. The overdot denotes differentiation with respect to the affine parameter $\lambda$.

Using Eqs.~\eqref{H_EG}-\eqref{eqm3_EG}, and $\mathrm{H}_{\pm} = 0$, which holds for photons, we obtain a radial equation given by
\begin{equation}
\label{RE_EG}\left(\dfrac{\dot{r}}{G^{\pm}_{1}(r)}\right)^{2} + \mathrm{V}_{\pm}(r) = E^{2},
\end{equation}
where $\mathrm{V}_{\pm}(r)$ is the effective potential for the radial motion of photons, defined as
\begin{equation}
\label{Veff_EG}\mathrm{V}_{\pm}(r) \equiv L^{2}\dfrac{G^{\pm}_{2}(r)f(r)}{G^{\pm}_{1}(r)r^{2}}.
\end{equation}

Closed circular photon orbits are described by $\dot{r} = 0$ and $\ddot{r} = 0$, which implies that 
\begin{equation}
\label{V12}\mathrm{V}_{\pm}(r)=E^2 \ \ \text{and} \ \ \mathrm{V}^{\prime}_{\pm}(r) = 0,
\end{equation}
respectively. Moreover, if $\mathrm{V}^{\prime\prime}_{\pm}(r)<0$, then the closed circular photon orbit is unstable. From $\dot{r} = 0$, we obtain the critical impact parameter $b_{l} = L_{l}/E_{l}$ associated with the LR, namely
\begin{equation}
\label{CIP_EG}b_{l} =  r_{l}\sqrt{\dfrac{G^{\pm}_{1}(r_{l})}{G^{\pm}_{2}(r_{l})f_{l}}},
\end{equation}
where $f(r_{l}) \equiv f_{l}$ and $r_l$ is the radial coordinate of the closed circular photon orbit.\footnote{We point out that the terminology ``light ring'' is widely used in the literature in a more general sense to denote circular photon orbits, independently of whether the spacetime is rotating or spherically symmetric~\cite{CH2018,PT2021}. In the present work, our use of the term follows this broader convention.} Moreover, from $\ddot{r} = 0$, we obtain the corresponding radial coordinate of the LR, given by
\begin{equation}
\label{CR_EG}f_{l}\left[2-\dfrac{r_{l}\left(G^{\pm}_{2}(r_{l})\right)^{\prime}}{G^{\pm}_{2}(r_{l})} + \dfrac{r_{l}\left(G^{\pm}_{1}(r_{l})\right)^{\prime}}{G^{\pm}_{1}(r_{l})} \right]-r_{l}f_{l}^{\prime} = 0,
\end{equation}
where the subscript ``$l$'' denotes that the quantity under consideration is computed at the radial coordinate of the LR $r_{l}$. Notice also that our results described above reduce to the RN case for $\mu = 0$, and to Schwarzschild for $Q = 0$.

\begin{figure}[!htbp]
\begin{centering}
    \includegraphics[width=\columnwidth]{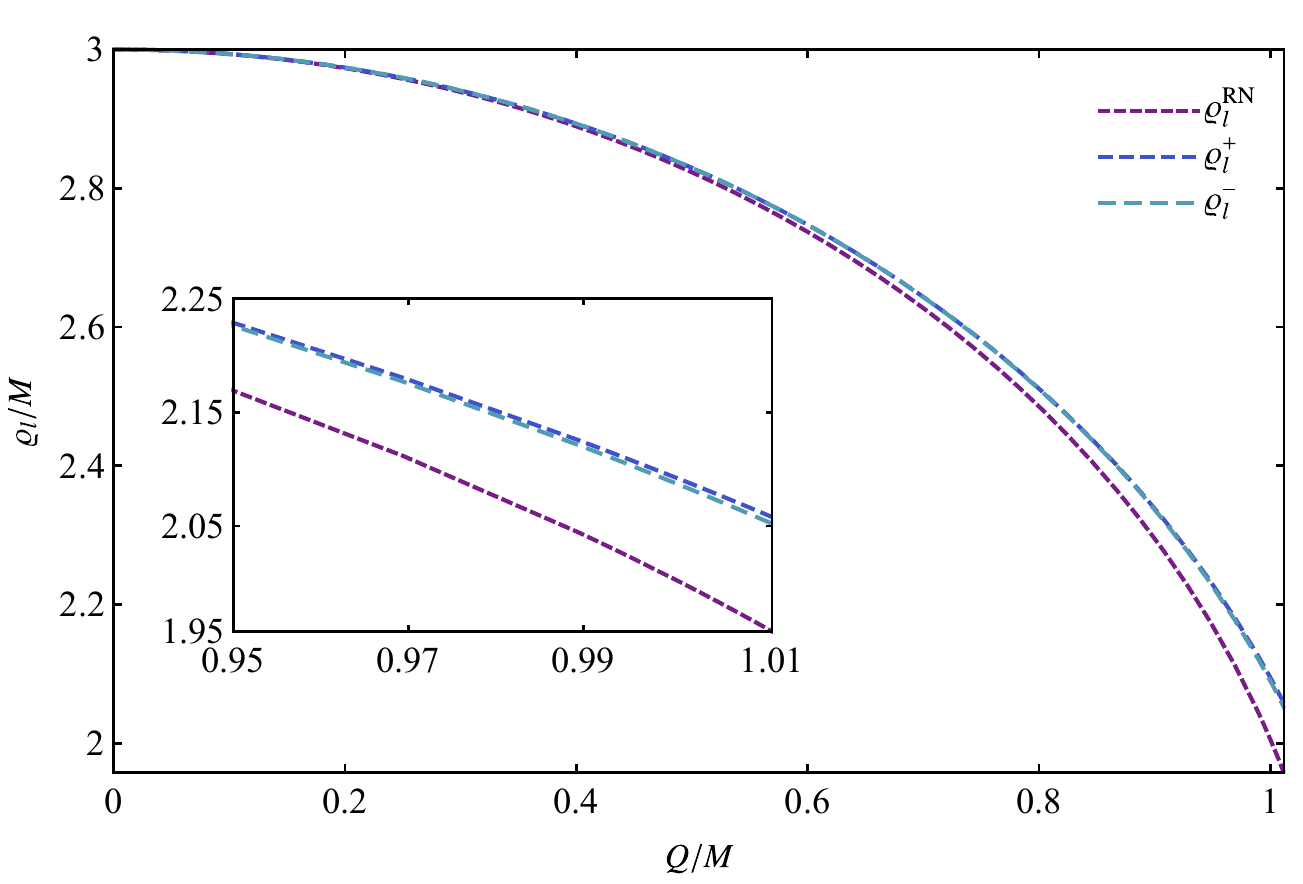}
    \caption{Comparison between the LRs of EH and RN BH geometries, as functions of $Q/M$. For the EH case, we consider the LRs of both polarizations. Here, we set $\mu = 0.05$, and $Q_{\rm{ext}} = 1.0115M$. The inset zoomed the LRs for extreme charge-to-mass BH values.}
    \label{lightrings}
\end{centering}
\end{figure}

In Fig.~\ref{lightrings}, we compare the LRs of the EH spacetime, considering the effective metrics, with that of the RN geometry. We consider the perimetral radius, defined by $\varrho \equiv \sqrt{g_{\varphi\varphi}}\big|_{\theta = \pi/2}$, which is a meaningful geometrical quantity to compare distances in different geometries. As we can see, the perimetral radius of the LRs for the EH spacetime is very similar for both polarizations of light. Moreover, they are usually larger than those in the RN case (recall that for the RN case $\varrho = r$).

\subsection{Analytical approximation for the radial coordinate of the LRs and critical impact parameter}
\label{Subsec:VB}
Considering that all physically relevant values of $\mu$ are much smaller than unity $(\mu \approx 0)$, the radii of the LRs and the critical impact parameter of the EH geometry resembles that of the RN spacetime with the same charge $Q$, as can be seen in Fig.~\ref{lightrings} and Fig.~\ref{nonasymobs} (see Sec. VI). Here, we derive an analytical approximation for the radii of the LRs in the limit $\mu \approx 0$. This analytical approximation is obtained using the Newton--Raphson method~\cite{PTV2007}, which determines the roots of a generic equation of the form
\begin{align}
R(r)=0,
\end{align}
through the following approximate expression
\begin{align}
r\approx r_0 -\frac{R(r_0)}{R'(r_0)},
\end{align}
where $r_0$ is an initial guess near the desired root. The equation for the radial coordinate of the LR is given by Eq.~\eqref{V12}, where one may identify $R(r)=\mathrm{V}^{\prime}_{\pm}(r)=0$, and apply the Newton-Raphson formula, obtaining
\begin{align}
r^{\pm}_l\approx r_0-\frac{\mathrm{V}^{\prime}_{\pm}(r_0)}{\mathrm{V}^{\prime\prime}_{\pm}(r_0)}.
\end{align}
In our case, a reasonable guess to $r_0$ is the radial coordinate of the LRs in the RN spacetime, obtained by setting $\mu=0$ in Eq.~\eqref{V12},
\begin{align} 
\label{NR_formula}r_0=\frac{3M}{2}+\sqrt{\frac{9}{4}M^2-2\,Q^2}.
\end{align}

Substituting $r_0$ into the Newton-Raphson formula~\eqref{NR_formula}, and considering $\mu\approx 0$, we obtain,
\begin{align}
\nonumber r^+_l&\approx\  r_0+\frac{\mu}{20Q^4\left(3M-2r_0\right)}\left[ 6M^2\left(29\,r_0^2-99Q^2\right)+ \right.\\
\label{NR_r+}&\left. \left(256MQ^2-783M^3\right)r_0+783M^4+132Q^4 \right]+\mathcal{O}\left[\mu^2\right],
\end{align}
for the $\mathcal{P}^+$ polarization, and 
\begin{align}
\nonumber r_l^-&\approx r_0+\frac{\mu}{10Q^4 \left(3M-2r_0\right)} \left[\left(68M Q^2-63M^3 \right)r_0+ \right.\\
\label{NR_r-}&\left. 189M^4-246 M^2 Q^2+36 Q^4 \right]+\mathcal{O}\left[\mu^2\right],
\end{align}
for the $\mathcal{P}^-$ polarization. 

\begin{figure}
\begin{centering}
    \includegraphics[width=\columnwidth]{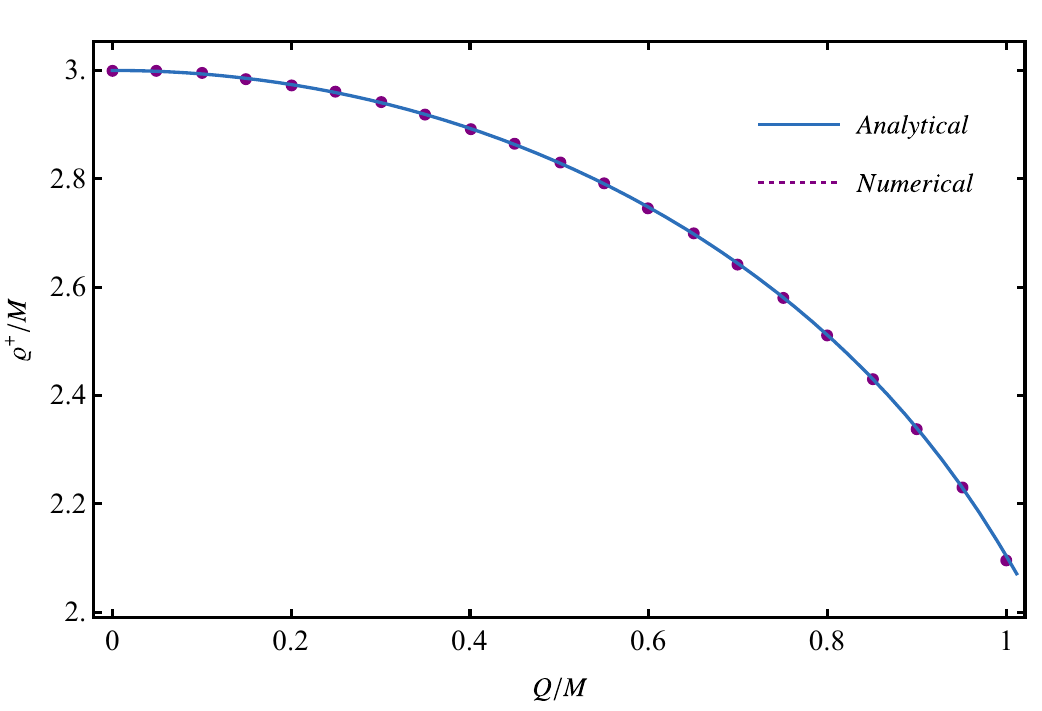}
	\includegraphics[width=\columnwidth]{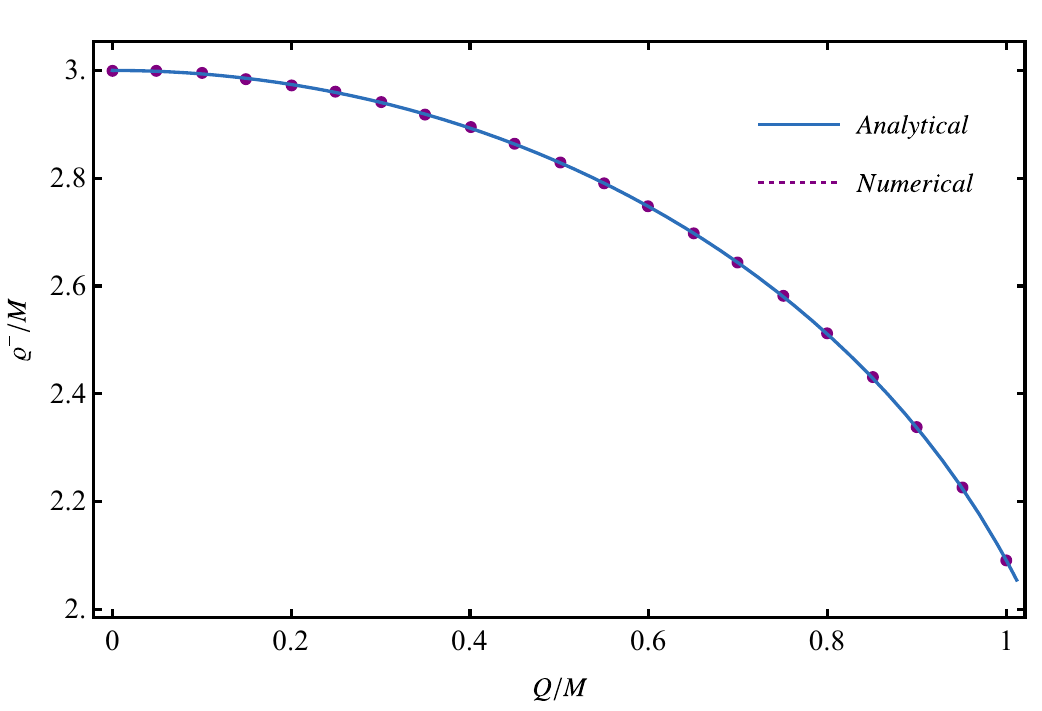}
    \caption{The comparison between the numerical results of Subsec.~\ref{Subsec:VA} and the analytical expressions for the radii of the LRs in the EH spacetime with $\mu=0.05$. In the top panel, we present the comparison for the $\mathcal{P}^+$ polarization, whereas the bottom panel shows the corresponding result for the $\mathcal{P}^-$ polarization.  }
    \label{comparison_an_num}
\end{centering}
\end{figure}

We plot the perimetral radius $\varrho$ associated with the LR radii for each polarization in Fig.~\ref{comparison_an_num}. In this figure, we show the comparison between the numerical results, discussed in Sec.~\ref{Subsec:VA}, and the analytical approximations~\eqref{NR_r+} and \eqref{NR_r-}, and notice that our analytical approximation describes very well the numerical results even for larger values of the charge $Q$. We also note that the linear-order contributions in $\mu$ appearing in Eqs.~\eqref{NR_r+} and \eqref{NR_r-} are strictly positive. This explains why the LRs radii of the EH geometry are always larger than those of the RN spacetime, in agreement with the numerical results shown in Fig.~\ref{lightrings}.

We can use the analytical expressions for $r_l^{\pm}$ to determine an analytical approximation for the critical impact parameter through Eq.~\eqref{CIP_EG}. Substituting Eqs.~\eqref{NR_r+} and \eqref{NR_r-} into Eq.~\eqref{CIP_EG}, and expanding for small values of $\mu$, we obtain
\begin{align}
\label{NR_b}b^{\pm}_l \approx b_0+b'^\pm_l \mu+\mathcal{O}\left[\mu^2\right],
\end{align}
where $b_0$ is the critical impact parameter of the RN spacetime,
\begin{align}
b_0=\frac{2\sqrt{2}\, Q\, r_0}{\sqrt{4Q^2-6M^2+2 M r_0}},
\end{align}
and $b'^\pm_l\equiv \left.\frac{d b^\pm_l}{d\mu}\right|_{\mu=0}$ is the first order coefficient of the expansion in $\mu$, given by
\begin{align}
&b'^\pm_l=\frac{2\,Q^3\left(\kappa_{1\pm}-\kappa_{2\pm} +68\,Q^2\,r_0\right)}{5r_0^3\left(2\,r_0-3M\right)(Q^2-M\,r_0)\sqrt{2Q^2+M(r_0-3M)}},
\end{align}
with
\begin{align}
&\kappa_{1+}=38M\,Q^2, \qquad \kappa_{2+}=105\,r_0M^2,\\
&\kappa_{1-}=53M\,Q^2, \qquad \kappa_{2-}=120\,r_0M^2, 
\end{align}
The $b'^\pm_l$ coefficient is strictly positive for all values of $Q/M$, implying that the critical impact parameter for both polarizations in the EH spacetime is always larger than that of the RN spacetime for small values of $\mu$. This analytical result agrees with our numerical results for the shadow of the EH BH, as can be seen in the right panel of Fig.~\ref{fpshadow} (see Sec.~\ref{sgl}), where the RN shadow is the smallest among all the cases. 

\subsection{LRs and topological charge in BHs sourced by NED}
\label{Subsec:VC}

In Sec.~\ref{Subsec:VA}, we obtained that the EH spacetime admits two LRs, one for each polarization, outside the event horizon. This result is remarkable when interpreted from the point of view of the topological charge associated with LRs, as introduced in Ref.~\cite{PV:2020}. In order to have a deeper understanding of this result, let us briefly review the main definitions of the topological charge associated with LRs.\footnote{For a more detailed discussion, we suggest the reader the Ref.~\cite{PV:2020}.}

LRs are critical points of the effective potential associated with the null geodesic motion. One can associate a topological charge to a LR by defining the vector field $\vec{v}=(v_r, v_\theta)$, as the normalized gradient of a (novel) effective potential, given by
\begin{align}
v_r=\frac{\partial_r U_{\pm}}{\sqrt{g_{rr}}}, \qquad v_\theta=\frac{\partial_\theta U_{\pm}}{\sqrt{g_{\theta\theta}}},
\end{align}
where 
\begin{align}
U_{\pm}\equiv\sqrt{\dfrac{G_2^\pm\,f(r)}{G_1^\pm\,r^2}}
\end{align}
is another effective potential, akin to the one introduced in Ref.~\cite{PV:2020}, which is independent of the constants of motion $E$ and $L$, but reproduce the same results as the effective potential $V_{\pm}(r)$, defined in Eq.~\eqref{Veff_EG}. At the coordinate of the LRs, the vector field $\vec{v}$ is null since LRs are critical points of the effective potential.

We can define an angle $\gamma$ that, together with the norm $v$, parametrizes the vector field as 
\begin{align}
\vec{v}=(v\,\cos\gamma, v\,\sin\gamma).
\end{align}
A topological quantity can be defined by computing the winding number of $\vec{v}$ when a closed curve $C$ is circulated in the anticlockwise sense,
\begin{align}
\oint_C d\gamma= 2\pi w,
\end{align}
where $w$ is an integer known as the winding number. LRs with a topological charge of $w=-1$ are referred to as standard LRs, with the unstable LRs being an example. In Ref.~\cite{PV:2020}, the following theorem was proved: \textit{``A stationary, axisymmetric, asymptotically flat black hole spacetime in 1+3 dimensions, with a nonextremal, topologically spherical, Killing horizon admits, at least, one standard LR outside the horizon for each rotation sense.''}

This result implies that a rotating BH satisfying all the hypotheses of the theorem must have at least one unstable LR for each sense of rotation, while any additional LRs must appear in stable/unstable pairs, in order to conserve the winding number $w$. In the case of a static BH, there must be at least one unstable LR, with any additional LRs also appearing in stable/unstable pairs.

Although this result is quite general, it does not account for the effects of two distinct effective metrics, as it is the case of BHs sourced by NED. Surprisingly, our findings indicate that such BHs can admit more than one LR for each sense of rotation or, in the case of a static BH, we obtained more than one unstable LR, one for each polarization, as shown in Fig.~\ref{lightrings}. The topological theorem proved in Ref.~\cite{PV:2020} can be generalized to include the vacuum birefringence phenomenon by defining one effective potential for each polarization. We will explore this generalization in a future work.

\section{Optical effects in the EH BH spacetime}
\label{sgl}

In this section, we discuss the observational setup consistent with each polarization of light and the relation between the vacuum birefringence and shadows. We also present our main results concerning the gravitational lensing of EH BHs, exploring, in particular, the role of the effective geometry. We also compare our results with the observational data.

\subsection{Shadow radius}

In order to obtain the shadow and gravitational lensing for the EH BH, it is necessary to discuss an observational setup consistent with the NED source, taking into account the effective metrics. To find this setup, we follow closely Ref.~\cite{MP2023} (see, in particular, Sec. IV A).

We consider the reference frame attached to a static observer, described by a set of four orthonormal vectors $\hat{\lambda}_{\hat{a}}^{\ \mu}$, dubbed as vierbein. The index with hat ranges from 0 to 3 and identifies the vectors in the vierbein. For the EH spacetime, the vierbein associated with a static observer is given by
\begin{align}
&\hat{\lambda}_{\hat{0}}^{\ \mu}=\left[ f^{-1/2}(r), 0 ,0, 0 \right],\\
&\hat{\lambda}_{\hat{1}}^{\ \mu}=\left[0, f^{1/2}(r) ,0, 0\right],\\
&\hat{\lambda}_{\hat{2}}^{\ \mu}= \left[0, 0, r^{-1},0 \right],\\
&\hat{\lambda}_{\hat{3}}^{\ \mu}=\left[0, 0, 0, {(r\sin\theta)^{-1}} \right].
\end{align}
The norm of the four-momentum, as measured by the static observer, is given by~\cite{MP2023}
\begin{equation}
\overline{p}_{\hat{a}}\bar{p}^{\hat{a}} = g^{\mu\nu}\bar{p}_{\mu}\overline{p}_{\nu},
\end{equation}
where $\bar{p}_{\hat{a}}$ are the components of the four-momentum of the photon projected into the vierbein $\hat{\lambda}^{\ \mu}_{\hat{a}}$, given by
\begin{equation}
\label{vierbain}\bar{p}_{\hat{a}} = \hat{\lambda}^{\ \mu}_{\hat{a}}\bar{p}_{\mu}.
\end{equation}

In the context of the EH geometry, the norm of the four-momentum, according to each polarization of the electromagnetic field, leads to
\begin{align}
\bar{p}_{\hat{a}}\bar{p}^{\hat{a}} & = -\dfrac{14\mu F^{\mu}_{\ \lambda}F^{\lambda \nu}}{1+5\mu F}\bar{p}_{\mu}\bar{p}_{\nu},\\
\bar{p}_{\hat{a}}\bar{p}^{\hat{a}} & = -\dfrac{8\mu F^{\mu}_{\ \lambda}F^{\lambda \nu}}{1-2\mu F}\bar{p}_{\mu}\bar{p}_{\nu},
\end{align}
where we used Eqs.~\eqref{gamma1} and~\eqref{gamma2}, and the fact that $\overline{g}^{\mu\nu}_{\pm}\overline{p}_\mu\overline{p}_\nu=0$. Also, by using the geodesic equations~\eqref{eqm1_EG}-\eqref{eqm3_EG}, we can rewrite the above equations, respectively, as
\begin{align}
\label{vectorpa}\bar{p}_{\hat{a}}\bar{p}^{\hat{a}} & = \dfrac{7\mu F L^{2}}{r^{2}\left(1+5\mu F\right)} \equiv \mathrm{\bar{p}}_{+},\\
\label{vectorpb}\bar{p}_{\hat{a}}\bar{p}^{\hat{a}} & = \dfrac{4\mu F L^{2}}{r^{2}\left(1-2\mu F\right)} \equiv \mathrm{\bar{p}}_{-},
\end{align}
where we define the functions $\mathrm{\bar{p}}_{\pm}$ for later convenience. Therefore, the norm of the four-momentum of photons with polarization $\mathcal{P}_+$ of the electromagnetic field is always null or spacelike with respect to the spacetime metric. In other words,
\begin{equation}
\mathrm{\bar{p}}_{+} \geq 0.
\end{equation}
On the other hand, for the polarization $\mathcal{P}_-$ of the electromagnetic field, we can also have timelike norm with respect to the standard geometry in the region defined by
\begin{equation}
r < r^{+}_{\rm{sig}},
\end{equation}
where $r^{+}_{\rm{sig}}$ is given by Eq.~\eqref{effrad}. Moreover, we note that, according to Fig.~\ref{effradius}, the radius $r^{+}_{\rm{sig}}$ is always located inside the event horizon. Hence, for the two possible polarizations of light, the norm of the four-momentum of photons are given by \textit{spacelike} curves outside the event horizon.

\begin{figure}[!htbp]
\begin{centering}
    \includegraphics[width=\columnwidth]{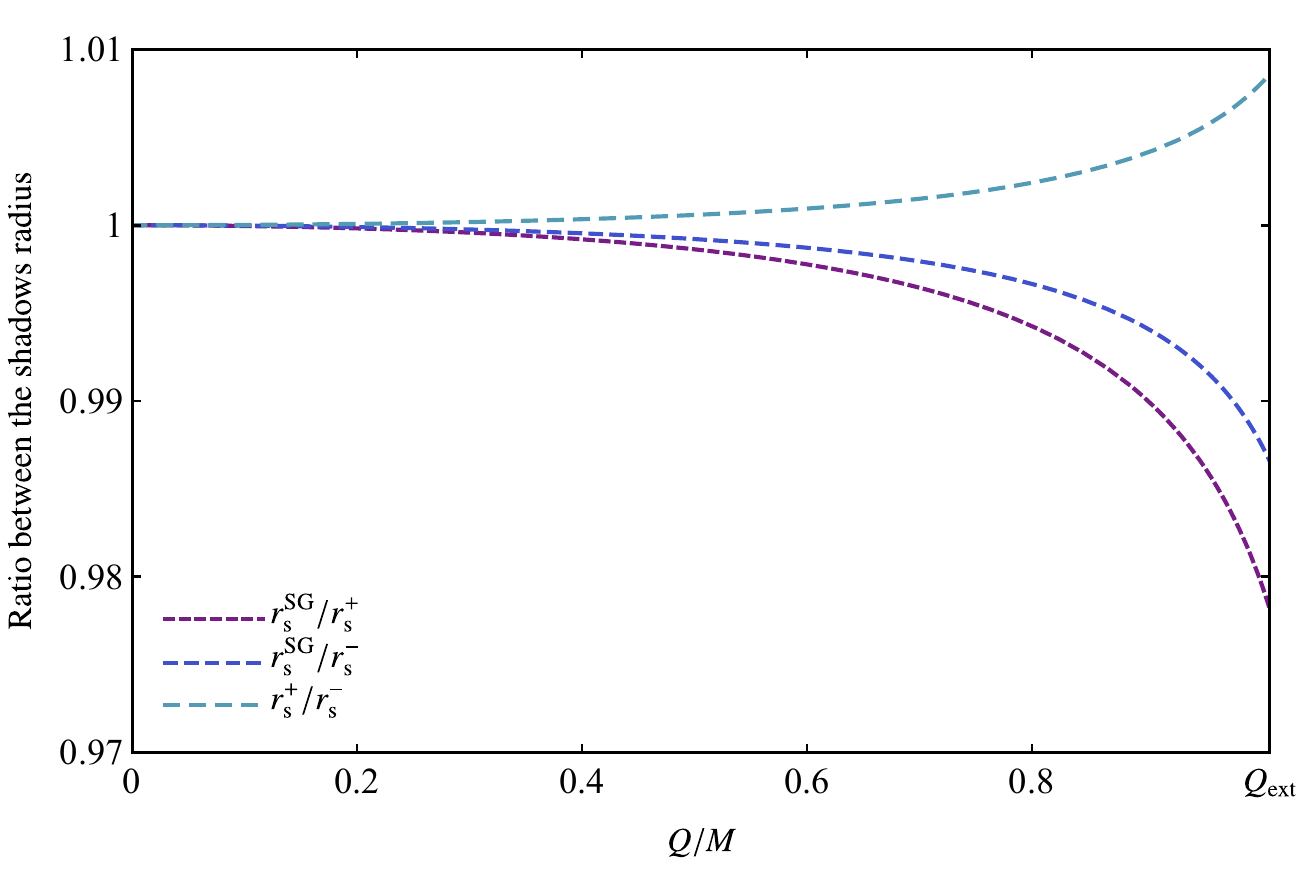}
	\includegraphics[width=\columnwidth]{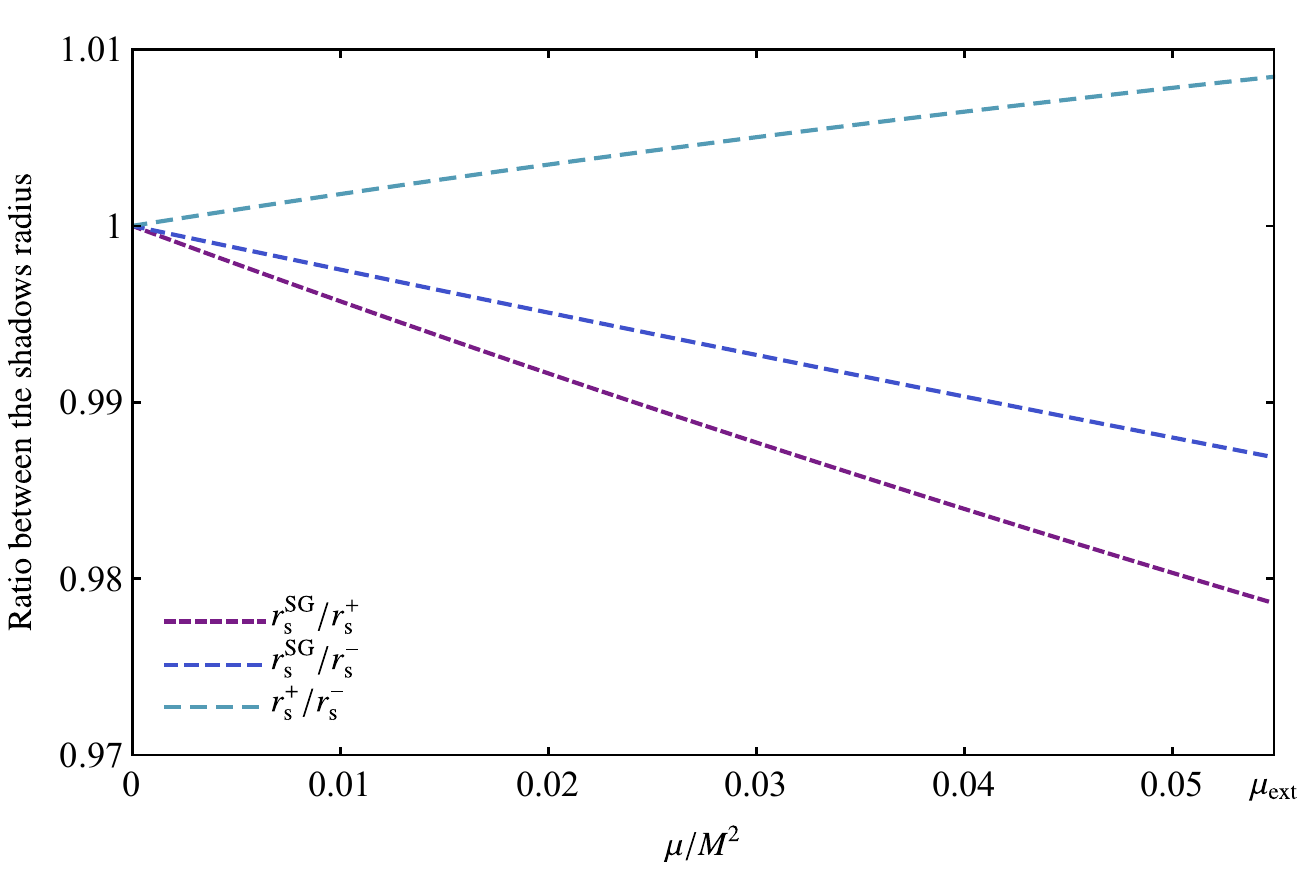}
    \caption{Ratio between the shadows radius of the effective geometry, $r_{s}^{\rm{SG}}$, and the effective metrics, $r_{s}^{\pm}$, as functions of $Q/M$ (top panel) and $\mu/M^{2}$ (bottom panel). In the top panel, we set $\mu/M^{2} = 0.05$, with $Q_{\rm{ext}} = 1.0115M$, while in the bottom panel, we fixed $Q = M$, with $\mu_{\rm{ext}} = 40M^{2}/729$.}
    \label{srr}
\end{centering}
\end{figure}

As also pointed out in Ref.~\cite{MP2023}, when the observer is placed at a finite radial coordinate outside the event horizon, the relation between the observational angles and the critical impact parameter is nontrivial. In what follows, we investigate the relation between the observational angles and the critical impact parameter in the context of the EH geometry. The spatial components of the four-momentum in terms of the celestial coordinates $(\alpha,\beta)$, defined in the reference frame of the static observer, can be written as:
\begin{align}
\label{pr}\bar{p}^{\hat{r}} & = \bar{\textbf{p}}\cos\alpha\cos\beta,\\
\label{ptheta}\bar{p}^{\hat{\theta}} & = \bar{\textbf{p}}\sin\alpha,\\
\label{pphi}\bar{p}^{\hat{\phi}} & = \bar{\textbf{p}}\cos\alpha\sin\beta,
\end{align}
where $\bar{\textbf{p}}$ is the norm of the photon's spatial three-momentum. Using Eqs.~\eqref{eqm1_EG}-\eqref{eqm3_EG},~\eqref{vierbain}, and~\eqref{pr}-\eqref{pphi}, we get
\begin{align}
\label{E_obs}E & = \bar{p}^{\hat{t}}\sqrt{f_{0}},\\
\label{rdot_obs}\dot{r} & = \bar{\textbf{p}}\sqrt{f_{0}}G^{\pm}_{1}(r_{0})\cos\alpha\cos\beta,\\
\label{thetadot_obs}\dot{\theta} & = \frac{\bar{\textbf{p}}\sin\alpha G^{\pm}_{2}(r_{0})}{r_0},\\
\label{L_obs}L & = \bar{\textbf{p}}\,r_0\,\sin\theta_0\cos\alpha\sin\beta,
\end{align}
where $(r_{0},\theta_{0})$ is the location of the observer and the subscript ``0'' denotes that the quantity under consideration is computed at the observer's position. Since photons follow spacelike curves outside the event horizon, with respect to the standard geometry, the relation between $\bar{p}^{\hat{t}}$ and $\bar{\textbf{p}}$ is not trivial. We can obtain the relation between these two quantities by considering Eqs.~\eqref{vectorpa} and \eqref{vectorpb}, and the fact that $\bar{p}_{\hat{a}}\bar{p}^{\hat{a}}=-(\bar{p}^{\hat{t}} )^2+ \bar{\textbf{p}}^2$. The result is given by
\begin{align}
\label{bart}&\overline{p}^{\hat{t}}=\left[1-\frac{14\mu Q^2}{\left(r_0^4+10\mu Q^2 \right)}\sin^2\beta\right]^\frac{1}{2} \overline{\textbf{p}}, \ \  (\text{for } \mathcal{P}_+),\\
\label{bartm}&\overline{p}^{\hat{t}}=\left[1-\frac{8\mu Q^2}{\left(r_0^4-4\mu Q^2 \right)}\sin^2\beta\right]^\frac{1}{2} \overline{\textbf{p}}, \ \ (\text{for } \mathcal{P}_-),
\end{align}
where we restricted our analysis to the equatorial plane ($\theta_0=\pi/2$ and $\alpha=0$). Dividing Eq.~\eqref{L_obs} by~\eqref{E_obs}, and using Eqs.~\eqref{bart}-\eqref{bartm}, we obtain the relation between the critical impact parameter $b_l$ and the observation angle of the shadow edge $\beta_l$, as measured by the local static observer, namely
\begin{align}
\label{obs_angle}\sin\beta_l=\dfrac{b_{l}\sqrt{f_{0}}}{r_{0}\sqrt{1+\bar{\mathrm{p}}_{\pm} b_{l}^{2}f_{0}}},
\end{align}
while the shadow radius yields
\begin{align}
\label{inf_obs}r_{s} = r_{0} \sin \beta_{l}.
\end{align}
If we place the observer far away from the central object, we get
\begin{equation}
\label{sr_NED}\lim_{r_{0} \rightarrow \infty}r_{s} = b_{l},
\end{equation}
with $b_{l}$ given by Eq.~\eqref{CIP_EG}. Hence, as seen by a distant observer, the impact parameter is the radius of the shadow, even in the presence of vacuum birefringence. In Fig.~\ref{srr}, we compare the shadow radius considering a distant observer in the standard geometry (i.e., described by $g_{\mu\nu}$) and also in the effective metrics of the magnetically charged EH BH (i.e., considering the polarizations $\mathcal{P}_+$ and $\mathcal{P}_-$). We notice that
\begin{equation}
\label{ratioshadows}r_{s}^{+} > r_{s}^{-} > r_{s}^{\rm{SG}},
\end{equation}
where $r_{s}^{+}$, $r_{s}^{-}$ and $r_{s}^{\rm{SG}}$ are the radius of the shadow computed using the polarization $\mathcal{P}_+$, the polarization $\mathcal{P}_-$, and the standard geometry, respectively. Therefore, the shadow radius computed with the effective metrics are larger than the shadow computed with the standard geometry.

We also point out that the shadow boundary for a distant observer can be expressed in terms of the so-called celestial coordinates $(x,y)$ as~\cite{B1973}
\begin{equation}
\label{sr_cc}r_{s} = \sqrt{x^{2}+y^{2}},
\end{equation}
where
\begin{align}
\label{cc1}x &= \lim_{r_{0} \rightarrow \infty}\left(-r_{0}\,\frac{\overline{p}^{\hat{\phi}}}{\overline{p}^{\hat{t}}} \right),\\
\label{cc2}y &= \lim_{r_{0} \rightarrow \infty}\left(r_{0}\,\frac{\overline{p}^{\hat{\theta}}}{\overline{p}^{\hat{t}}} \right),
\end{align}
and the shape of the shadow edge can be obtained from a parametric plot of the circle equation~\eqref{sr_cc}.

\begin{figure}[h!]
\begin{centering}
    \includegraphics[width=\columnwidth]{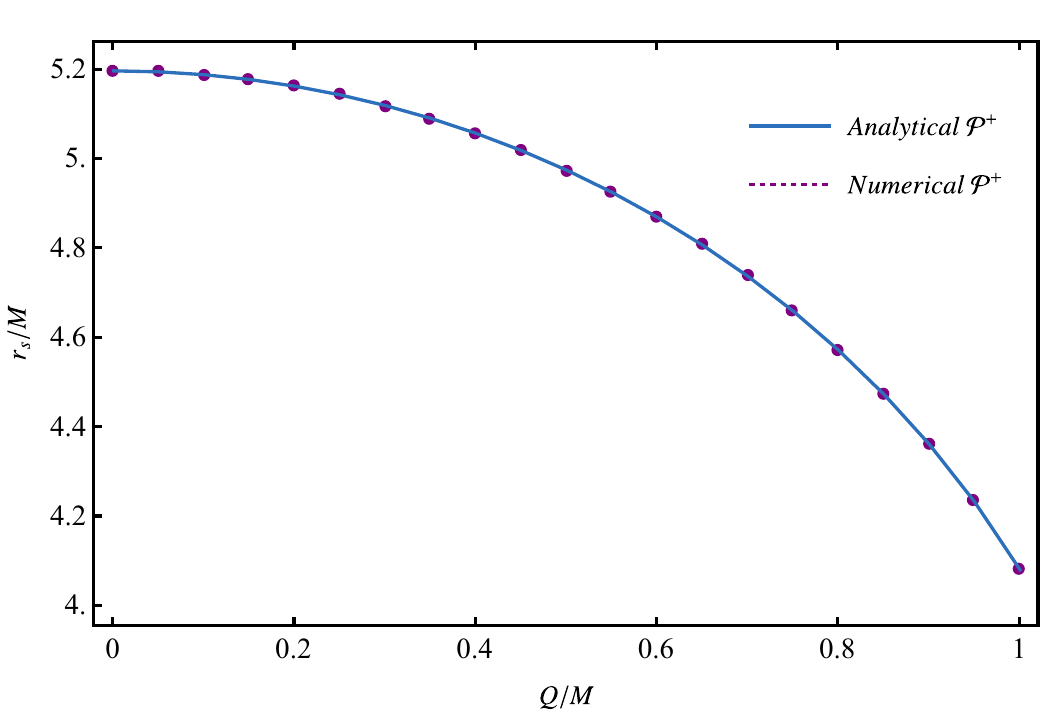}
	\includegraphics[width=\columnwidth]{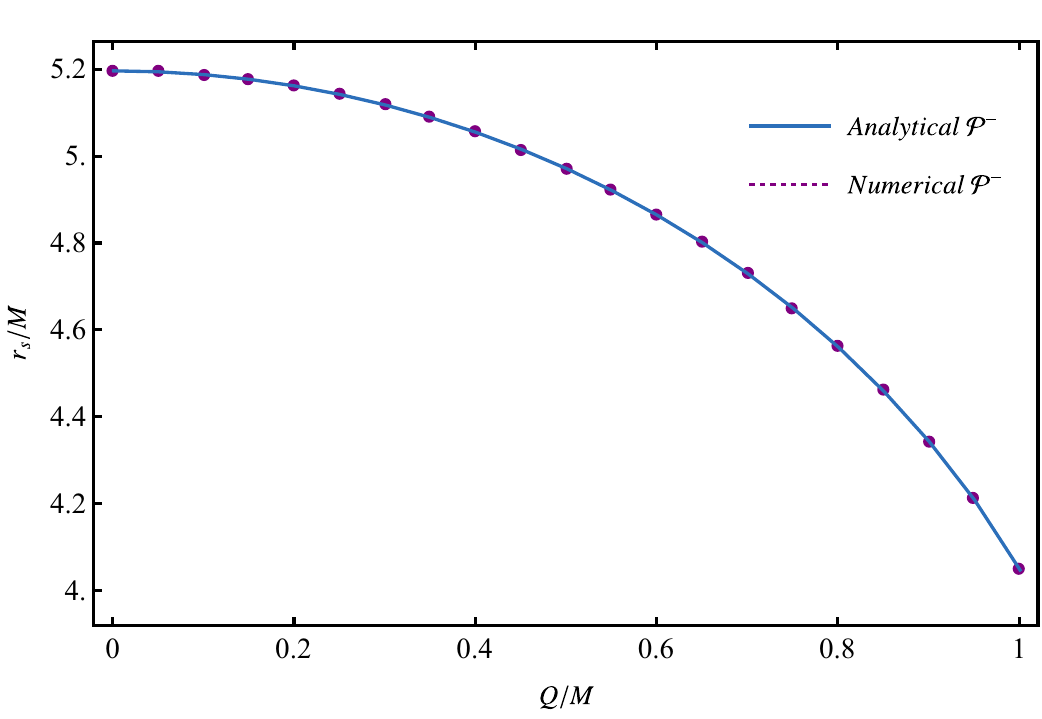}
    \caption{The comparison between the numerical results and the analytical approximation for the shadow radii in the EH spacetime with $\mu/M^{2}=0.05$. In the top panel, we present the shadow edge for the $\mathcal{P}^+$ polarization, whereas the bottom panel shows the corresponding result for the $\mathcal{P}^-$ polarization.  }
    \label{comparison_shadow_an_num}
\end{centering}
\end{figure}

\begin{figure}[h!]
\begin{centering}
    \includegraphics[width=\columnwidth]{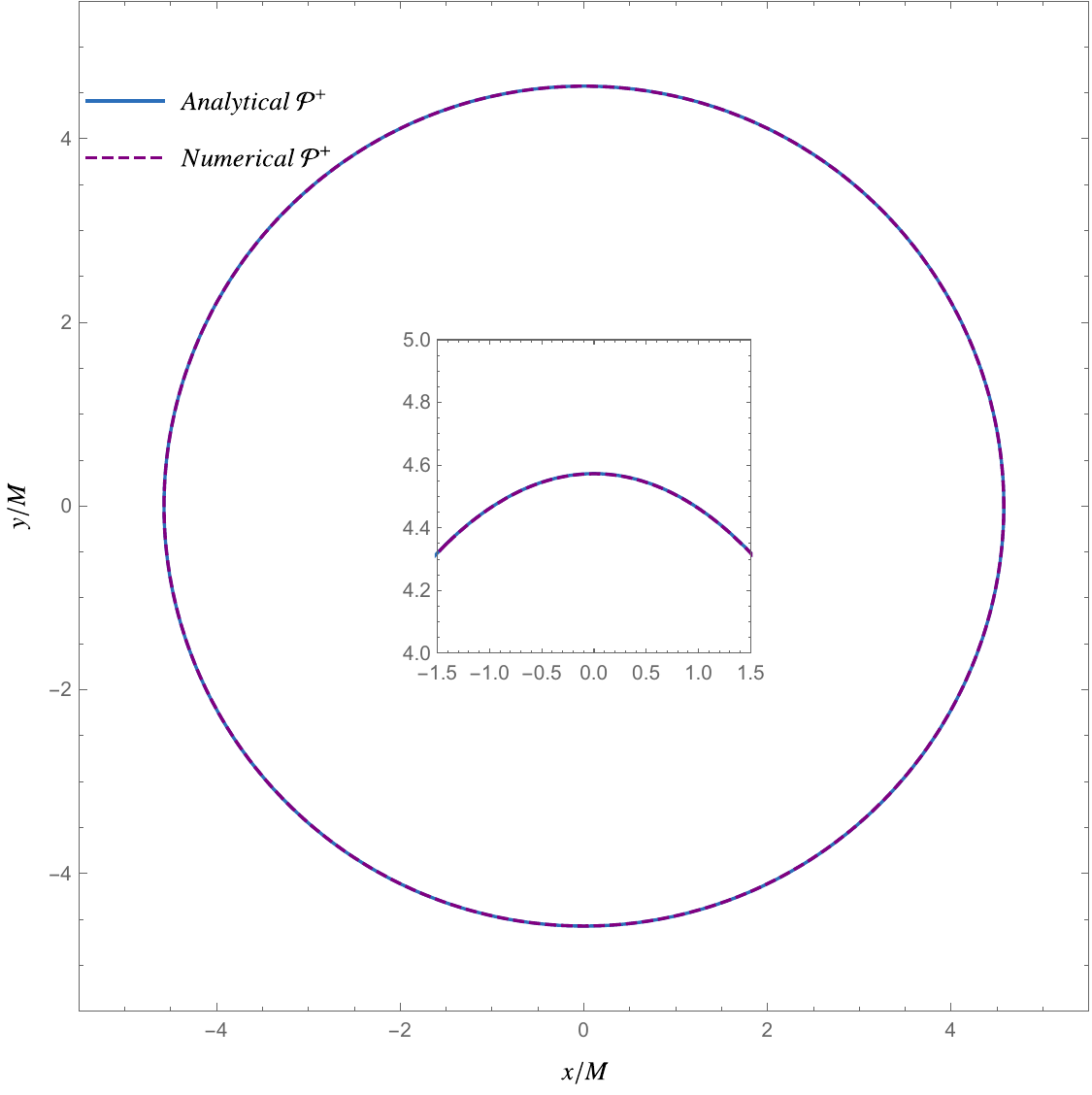}
    \caption{The comparison between the numerical result and the analytical approximation for the shadow edge of the $\mathcal{P}^+$ polarization, with $\mu/M^{2}=0.05$ and fixed charge $Q/M=1$.}
    \label{comparison_shadow_an_num2}
\end{centering}
\end{figure}

In Fig.~\ref{comparison_shadow_an_num}, we show the comparison between the numerical result for the shadow edge and the analytical approximation, given by Eq.~\eqref{NR_b}. As expected for $\mu\approx 0$, the analytical approximation agrees very well with the numerical result. Moreover, the analytical approximation allows us to draw general conclusions regarding the shadow edge, such as the property that the shadow of the EH BH is always larger than the RN BH, when both spacetimes have the same value of $Q/M$. In Fig.~\ref{comparison_shadow_an_num2}, we show the shadow edge of the EH BH computed for the $\mathcal{P}^+$ polarization, comparing the numerical result with the analytical approximation. As can be seen, the two shadow contours are practically indistinguishable. A similar result is obtained for the $\mathcal{P}^-$ polarization. 

\begin{figure*}[!htbp]
\begin{centering}
    \includegraphics[width=\columnwidth]{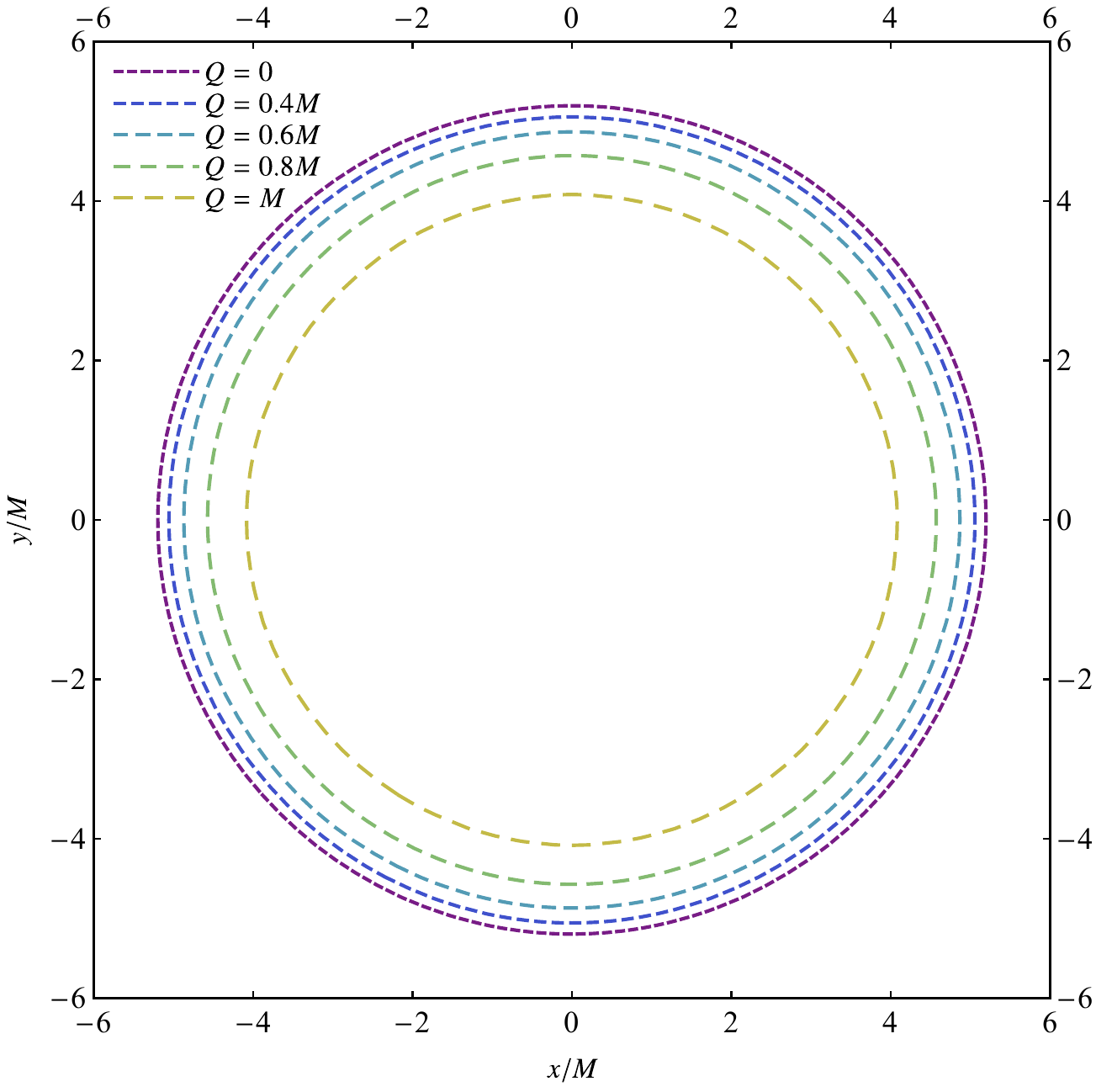}
	\includegraphics[width=\columnwidth]{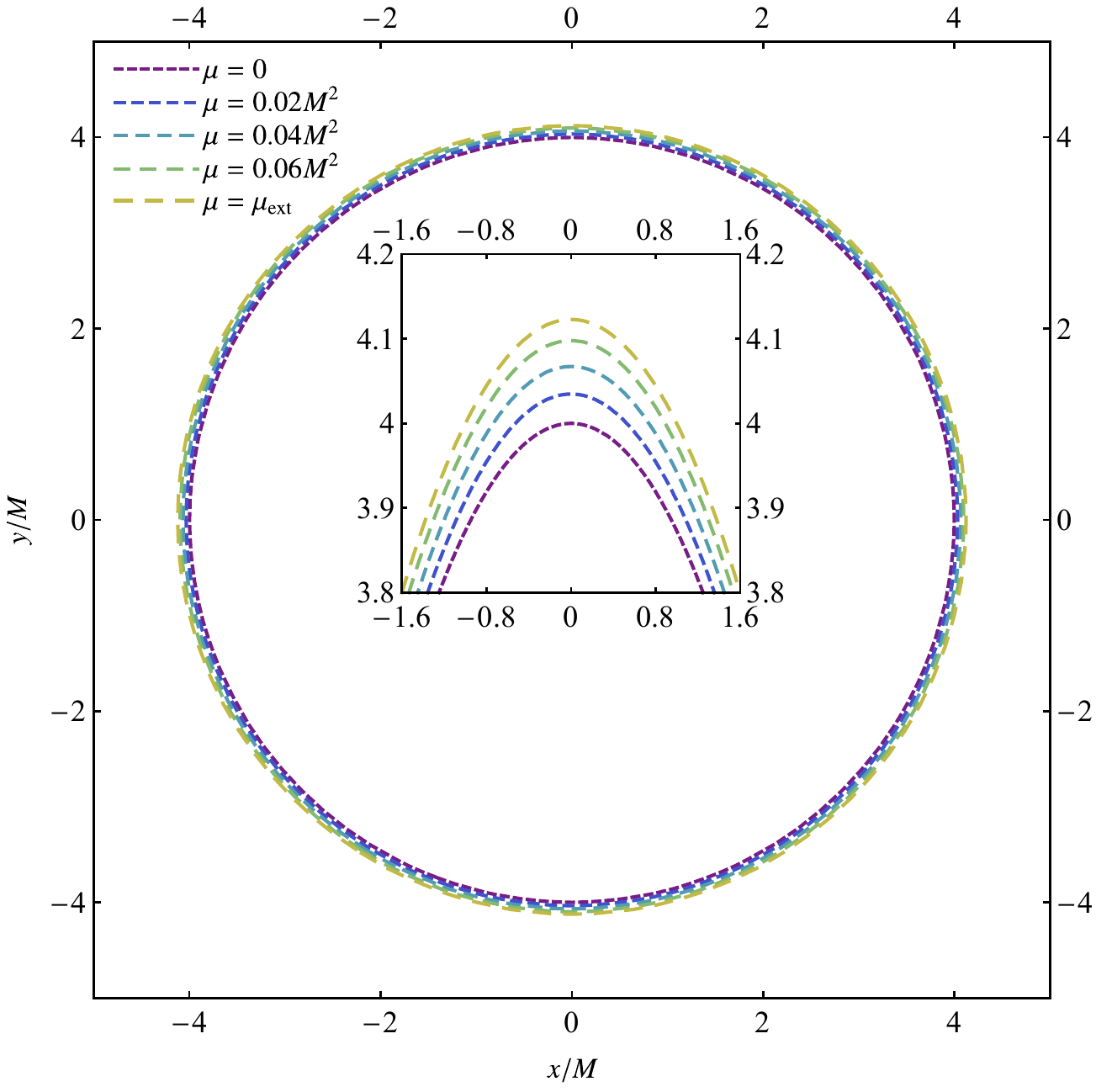}
    \caption{Shadows edge in the $(x, y)$-plane for different values of $Q/M$ (left panel) and $\mu/M^{2}$ (right panel), considering the effective metric associated with the first polarization of the electromagnetic field. In the first panel, we set $\mu = 0.05M^{2}$, while in the second panel, we fixed $Q/M = 1$. For comparison, we also exhibit the Schwarzschild and RN cases in each panel, respectively. The inset in the right panel helps us to better visualize the shadows edge for different values of $\mu/M^{2}$ in a given range of the $(x, y)$-plane.}
    \label{fpshadow}
\end{centering}
\end{figure*}

\begin{figure}[!htbp]
\begin{centering}
    \includegraphics[width=\columnwidth]{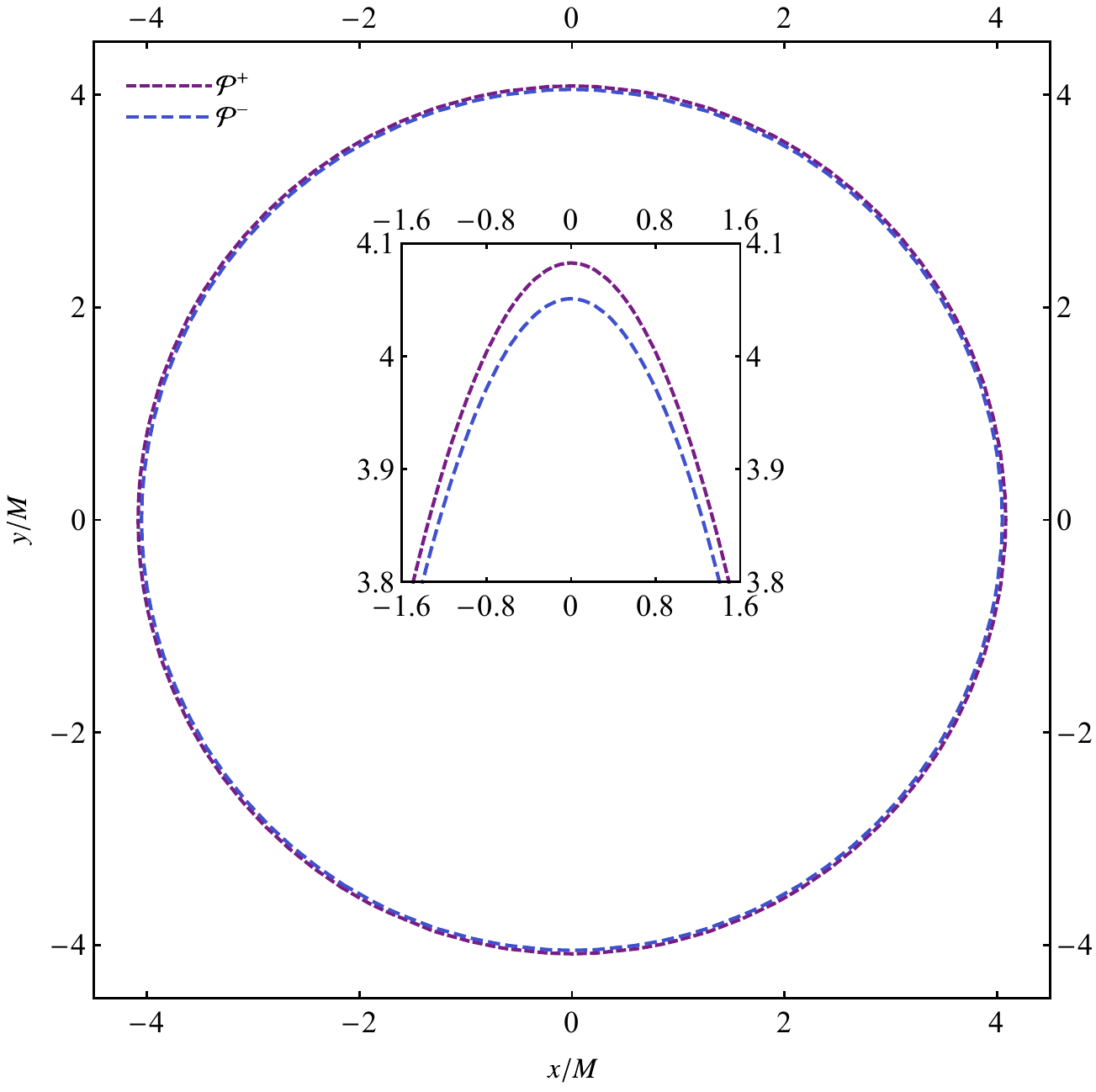}
    \caption{Comparison between the shadow edge in the $(x, y)$-plane of the first and second polarizations of the magnetic field for $Q/M = 1$ and $\mu = 0.05M^{2}$. The inset helps us to better visualize the shadows edge in a given range of the $(x, y)$-plane.}
    \label{fpandspsameQ}
\end{centering}
\end{figure}

In Fig.~\ref{fpshadow}, we display the shadow edge of the EH BH, computed with the polarization $\mathcal{P}_+$, considering an asymptotic observer. As we can see from Fig.~\ref{fpshadow}, the shadow edge decreases (increases) as we consider higher values of $Q/M$ $(\mu/M^{2})$. Consequently, we conclude that the critical impact parameters have a qualitatively similar behavior.
The behavior for the polarization $\mathcal{P}_-$ is quantitatively very similar to the polarization $\mathcal{P}_+$, as can be seen in Fig.~\ref{fpandspsameQ}, with $r_{s}^{+} > r_{s}^{-}$, as expected from Eq.~\eqref{ratioshadows}. Our results also indicate that the polarization $\mathcal{P}_+$ exhibits a pattern similar to Fig.~\ref{fpshadow} for different values of $Q/M$ or $\mu/M^{2}$.

\begin{figure}[!htbp]
\begin{centering}
    \includegraphics[width=\columnwidth]{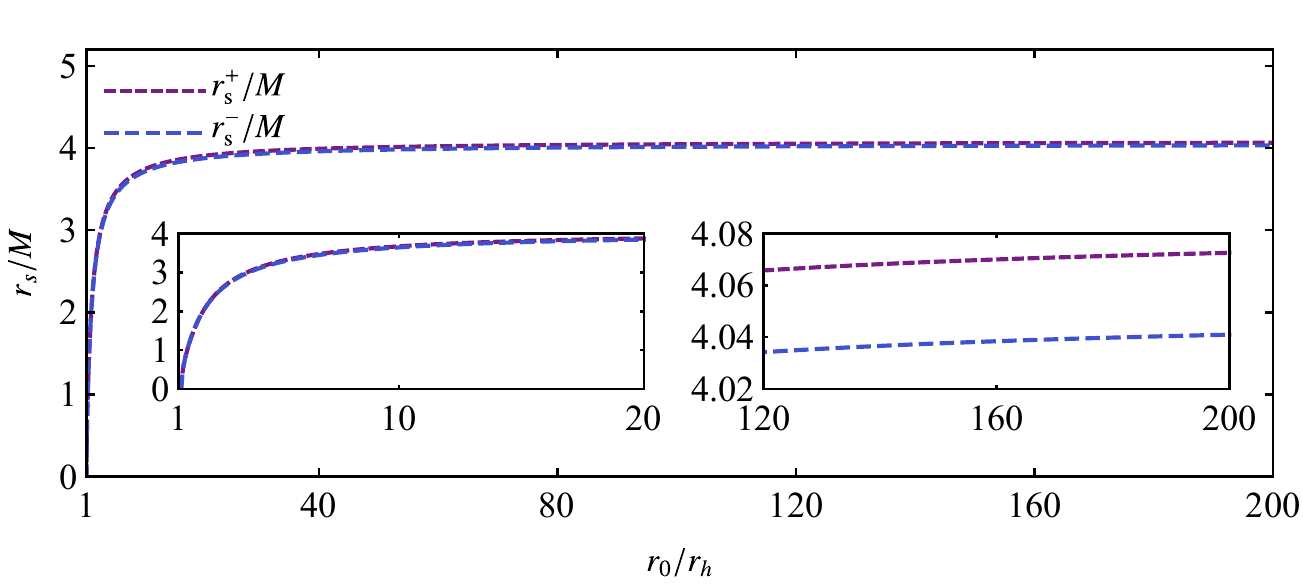}
    \caption{Shadows radii as functions of observer's position, normalized by the event horizon $(r_{0}/r_{h})$. In the insets, we exhibit the shadows radii near the horizon and close to the observer position, which we choose to be $r_{0} = 200r_{h}$. Here we set $\mu = 0.05M^{2}$ and $Q = M$.}
    \label{nonasymobs}
\end{centering}
\end{figure}

It is also interesting to investigate the role of the observer's position $r_{0}$ in the shadow radius, as presented in Fig.~\ref{nonasymobs}. We notice that the behavior of $r_{s}^{+}$ and $r_{s}^{-}$ is qualitatively similar as we increase $r_{0}$, and far from the BH the result given by Eq.~\eqref{sr_NED} is recovered. Notably, the shadows radii are strongly affected by the position of the observer near the horizon. This can be attributed to the strength of the magnetic field in the near horizon region.

\subsection{Gravitational lensing}

We can investigate gravitational lensing around EH BHs using the so-called backwards ray-tracing technique. This method is widely used in the literature to study the shadow and gravitational lensing of BHs, wormholes, and other compact objects. The backwards ray-tracing technique allows the simulation of a BH image by numerically solving the geodesic equations, tracing light rays backward in time from the observer's position until they reach either the event horizon or a distant celestial sphere concentric with the BH. By assigning different colors (e.g., red, green, blue, and yellow) to this celestial sphere, we can map the direction in which a given light ray was scattered by the BH.
\begin{figure}[!h]
\begin{centering}
    \includegraphics[width=\columnwidth]{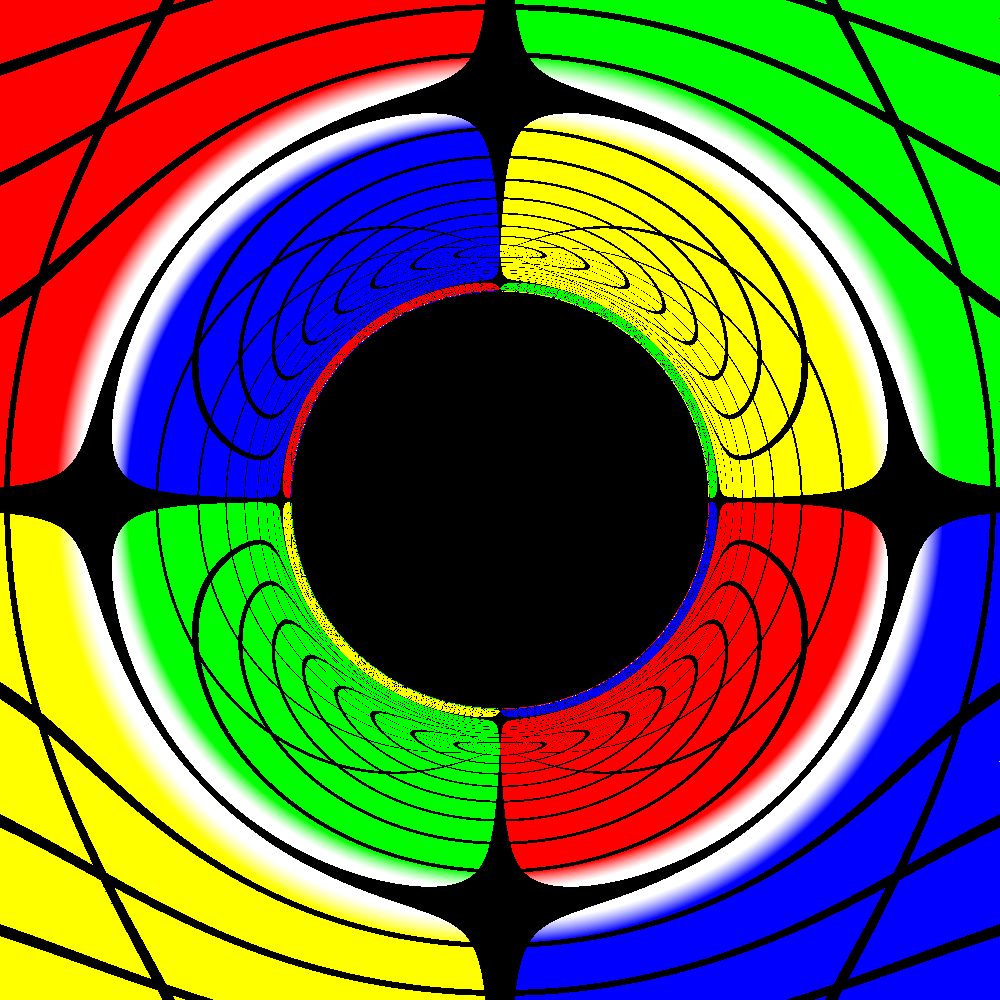}
	\includegraphics[width=\columnwidth]{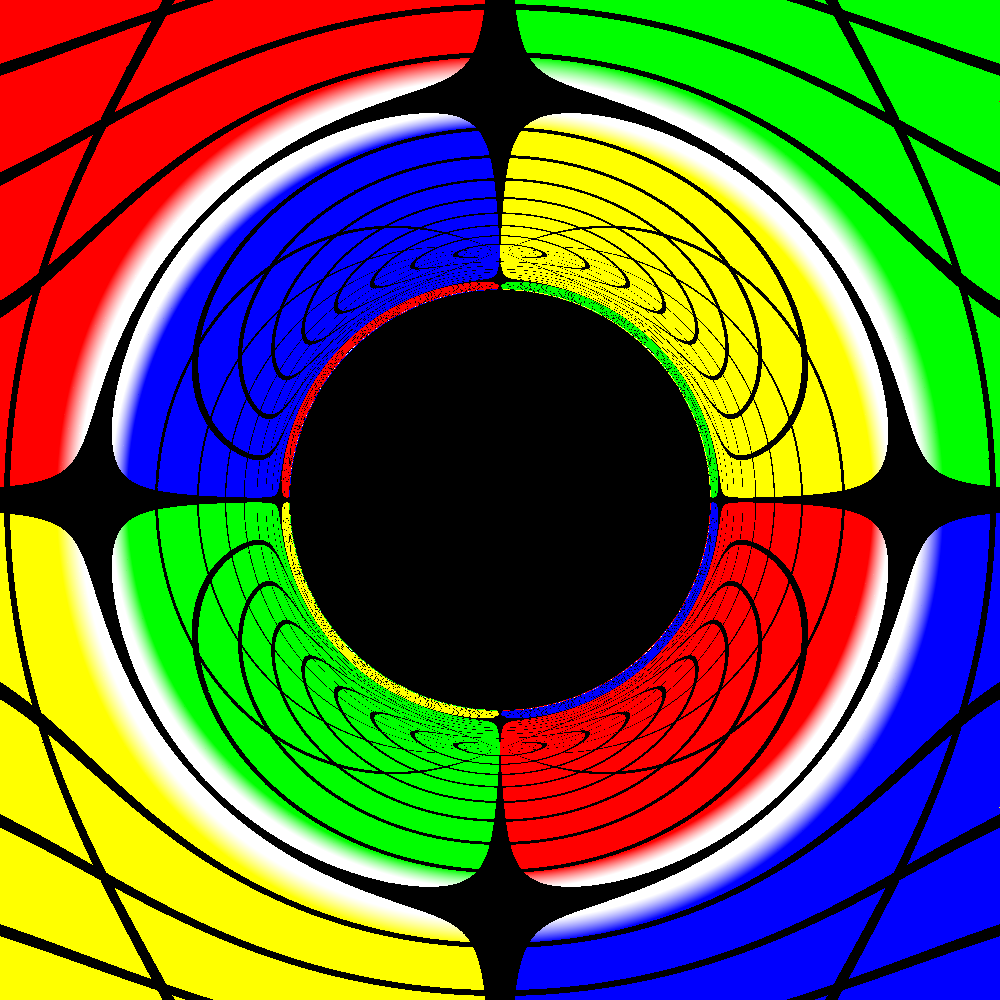}
    \caption{Backwards ray-tracing images of the shadow and gravitational lensing of EH BHs with $Q=M$ and $\mu=0.05M^2$. The top panel displays the result for the polarization $\mathcal{P}_-$, while the bottom panel corresponds to the  $\mathcal{P}_+$ polarization. }
    \label{Ray_tracing_Q1_mu005}
\end{centering}
\end{figure}

\begin{figure}[!h]
\begin{centering}
    \includegraphics[width=\columnwidth]{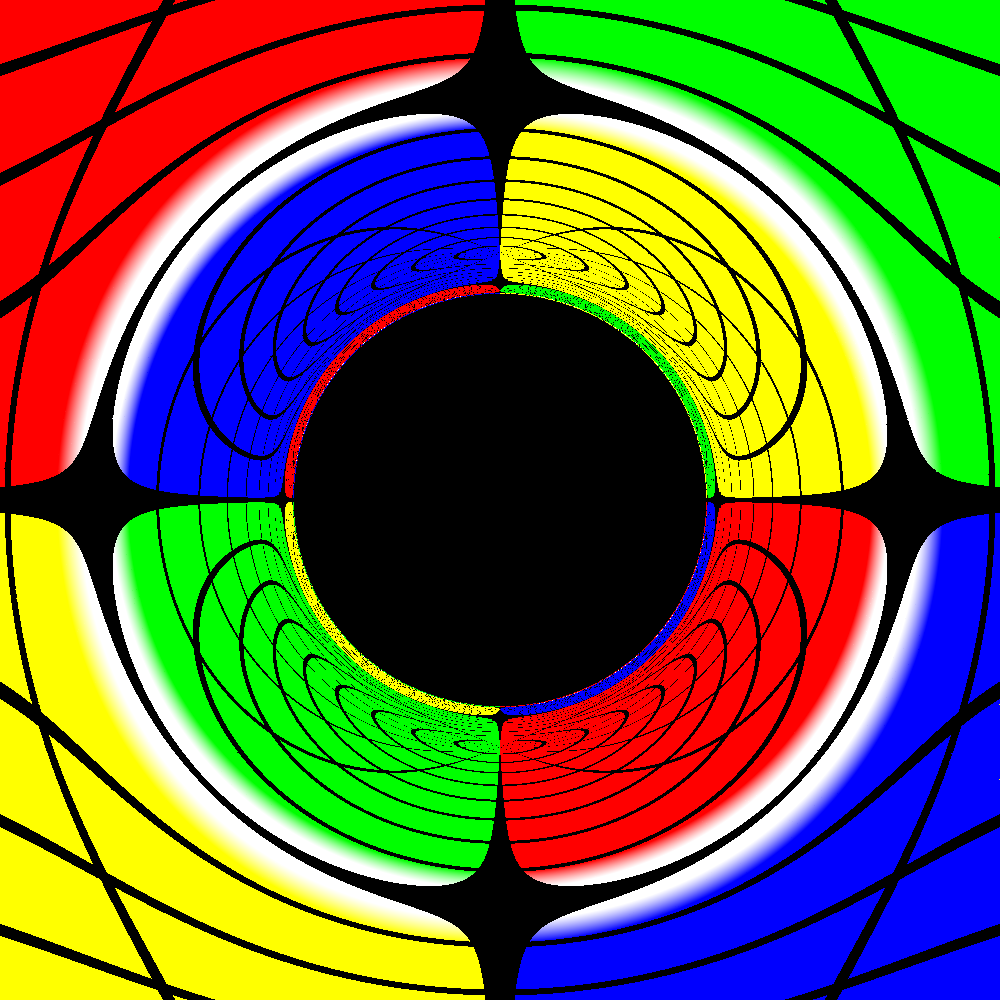}
	\includegraphics[width=\columnwidth]{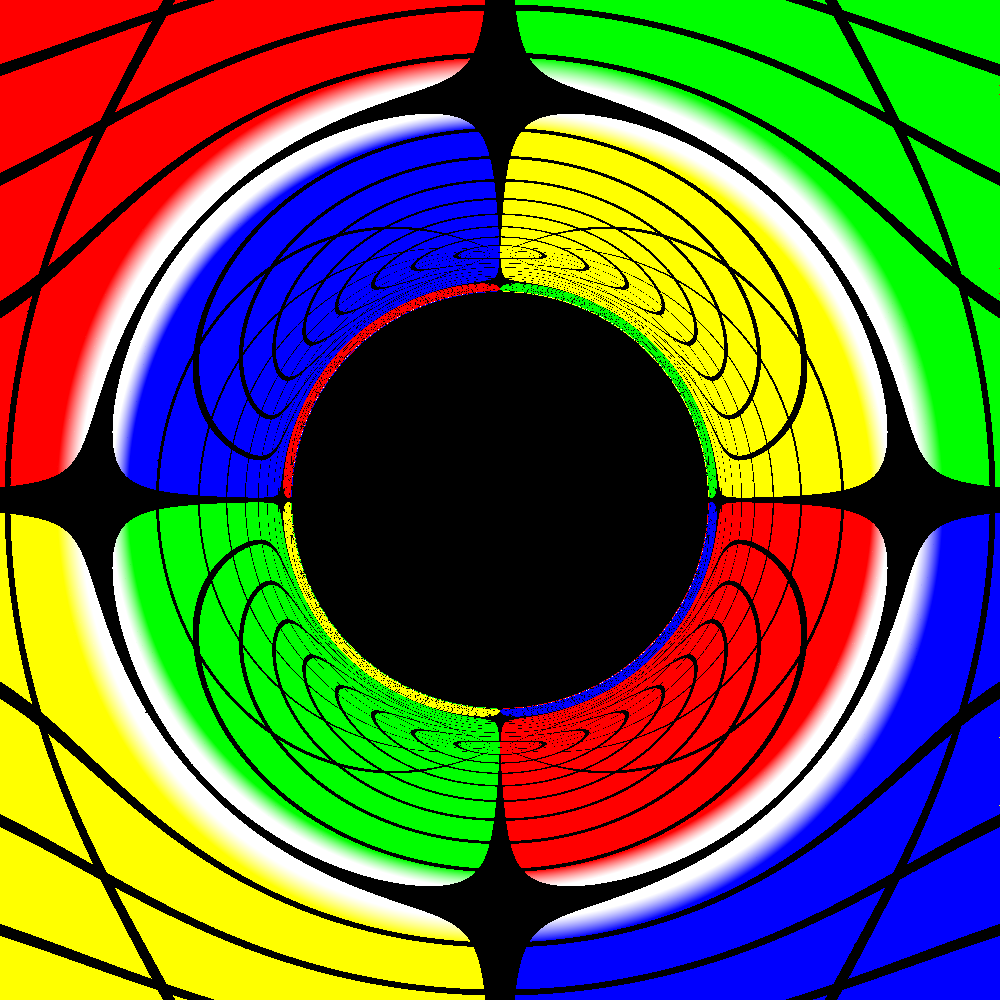}
    \caption{Backwards ray-tracing images of the shadow and gravitational lensing of extreme EH BHs with $Q=1.0115M (=Q_{ext})$ and $\mu=0.05M^2$. The top panel displays the result for the polarization $\mathcal{P}_-$, while the bottom panel corresponds to the $\mathcal{P}_+$ polarization.}
     \label{Ray_tracing_Qext_mu005}
\end{centering}
\end{figure}

The implementation of the backwards ray-tracing is based on solving the second order geodesic equations for $r$ and $\theta$:
\begin{align}
\label{ddotr}&\ddot{r}+\left(\overline{\Gamma}^r_{\ \mu\nu}\right)_{\pm} \dot{x}^\mu \dot{x}^\nu=0,\\
\label{ddottheta}&\ddot{\theta}+\left(\overline{\Gamma}^{\theta}_{\ \mu\nu}\right)_{\pm} \dot{x}^\mu \dot{x}^\nu=0,
\end{align}
coupled to the first-order equations for $t$ and $\phi$ given in Eqs.~\eqref{eqm1_EG} and \eqref{eqm3_EG}. In Eqs.~\eqref{ddotr}-\eqref{ddottheta}, $\left(\overline{\Gamma}^\alpha_{\ \mu\nu}\right)_{+}$ and $\left(\overline{\Gamma}^\alpha_{\ \mu\nu}\right)_{-}$ denote the Christoffel symbols computed with the effective metrics $\bar{g}^{\mu\nu}_{+}$ and $ \bar{g}^{\mu\nu}_{-}$, respectively. The Christoffel symbols as a function of the effective metrics (and their derivatives) are given by
\begin{align}
\left(\overline{\Gamma}^\mu_{\ \alpha\beta}\right)_{\pm} = \dfrac{\bar{g}^{\mu\rho}_{\pm}}{2}\left(\partial_\alpha \left(\bar{g}_{\rho\beta}\right)_{\pm}+\partial_\beta \left(\bar{g}_{\alpha\rho}\right)_{\pm}-\partial_\rho \left(\bar{g}_{\alpha\beta}\right)_{\pm} \right).
\end{align}
From Eqs.~\eqref{ddotr}-\eqref{ddottheta}, a light ray can follow one of two possible paths, described either by $\bar{g}^{\mu\nu}_{+}$ or $ \bar{g}^{\mu\nu}_{-}$,  depending on its polarization. In what follows, we solve the geodesic equations for both polarizations. In order to numerically solve Eqs.~\eqref{eqm1_EG}, \eqref{eqm2_EG}, \eqref{ddotr}, and \eqref{ddottheta}, we use as initial conditions,
\begin{align}
t(0)=0,\quad r(0)=r_0, \quad \theta(0)=\theta_0,\quad \phi(0)=0,
\end{align}
corresponding to the observer's position where the light rays originate. The remaining initial conditions are given by Eqs.~\eqref{E_obs}-\eqref{L_obs}, corresponding to the initial direction that light rays are traced backwards. The initial direction of light rays is parametrized in terms of the observational angles $\left(\beta, \alpha \right)$.

We can construct an image where to each pixel is assigned a color based on the light ray's trajectory: black if the ray terminates at the event horizon, or one of four distinct colors corresponding to the celestial sphere if the ray is scattered. We assume the same color pattern for the celestial sphere as used in Refs.~\cite{Bohn:2015, Cunha:2015, HCD:2021}. Repeating the process of numerically evolving the geodesic equations for a sufficient number of observational angles $\left(\beta, \alpha \right)$, we create the image simulating the visual appearance of an EH BH with the desired quality. In the remainder of this section, we present images with resolution $1000\times 1000$ pixels, obtained from the numerical evolution of $10^6$ geodesics. These images were generated using a C\texttt{++} code that employs the Dormand-Prince method to numerically solve the geodesic equations. This code has been used to simulate the image of other BH models in Refs.~\cite{MAP2024d,HCD:2021}. We also double-checked our results using the PyHole Python package, see Ref.~\cite{PVP2016}, in particular Appendix D, for more details.

In Fig.~\ref{Ray_tracing_Q1_mu005} we show the shadow and gravitational lensing of an EH BH with $Q=M$ and $\mu=0.05\,M^2$. In the top panel, the backwards ray-tracing for the polarization $\mathcal{P}_+$ is presented, while in the bottom panel we show the result for the $\mathcal{P}_-$ polarization. In Fig.~\ref{Ray_tracing_Qext_mu005}, we present the shadow and gravitational lensing of an EH BH with $Q=1.0115M$ and $\mu=0.05M^2$, corresponding to the extreme BH case. We also compare the backward ray-tracing images obtained for the $\mathcal{P}_+$ (top panel) and $\mathcal{P}_-$ (bottom panel) polarizations. We notice that in both Figs.~\ref{Ray_tracing_Q1_mu005} and~\ref{Ray_tracing_Qext_mu005} the images exhibit similar qualitative aspects, namely a central black region representing the BH shadow and the gravitationally lensed image of the colored celestial sphere. The white ring in the image represents the gravitationally lensed image of a white spot located on the celestial sphere. This white circle on the image plane corresponds to the well-known \textit{Einstein ring}~\cite{Einstein:1936}.

Although the backward ray-tracing images in the top and bottom panels of Figs.~\ref{Ray_tracing_Q1_mu005} and \ref{Ray_tracing_Qext_mu005} appear to be the same, they exhibit subtle differences for the distinct polarizations. To analyze the differences in the images of EH BHs for different polarizations, we compute their pixel-wise difference. Each pixel in the backward ray-tracing images can be converted to a grayscale,  with an intensity value $I(\beta, \alpha)$ ranging from zero to one. In order to compute the difference between the images obtained with polarizations $\mathcal{P}_+$ and $\mathcal{P}_-$, keeping fixed the values of $Q$ and $\mu$, we compute the subtraction of the intensities of the two images,
\begin{align}
\delta I(\beta, \alpha)= |I_{\mathcal{P}_+}(\beta, \alpha)-I_{\mathcal{P}_-}(\beta, \alpha) |,
\end{align}
where $I_{\mathcal{P}_+}$ ($I_{\mathcal{P}_-}$) denotes the intensity of the $\mathcal{P}_+$ ($\mathcal{P}_-$) polarization image. Thus, we can generate a grayscale image representing the difference between the two polarizations images of a single EH BH.
 
\begin{figure}[!h]
\begin{centering}
    \includegraphics[width=\columnwidth]{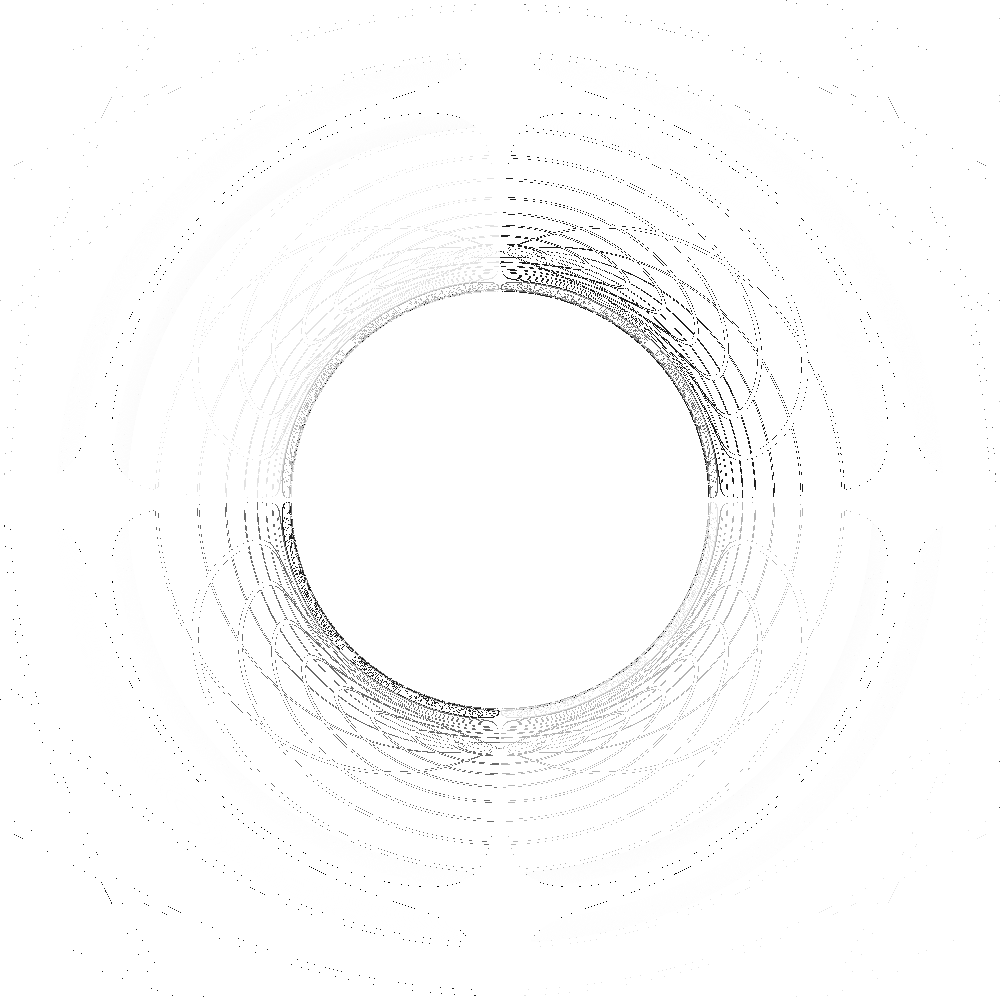}
	\includegraphics[width=\columnwidth]{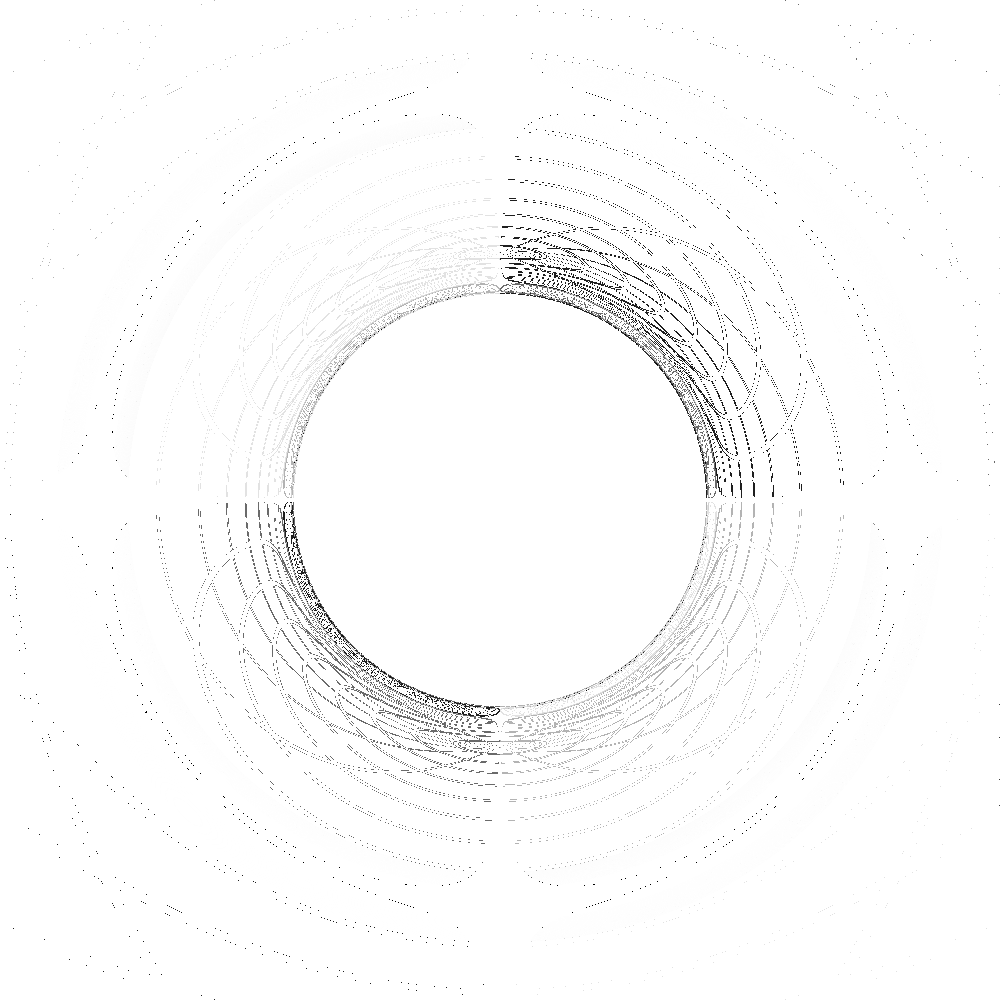}
    \caption{Gray scale difference image comparing the backwards ray-tracing for the two possible polarizations in the EH spacetime. The top panel represents the case of an EH BH with $Q=M$ and $\mu=0.05\,M^2$, while the bottom panel corresponds to the extreme EH spacetime with  $Q=1.0115M$ and $\mu=0.05\,M^2$. From this figure, we conclude that there are subtle differences between the images obtained with the two different polarizations. }
    \label{Difference_Image}
\end{centering}
\end{figure}

In Fig.~\ref{Difference_Image}, we present the gray scale difference image between the two polarizations of the EH BH. In the top panel of Fig.~\ref{Difference_Image}, we present the difference image between the polarizations $\mathcal{P}_+$ and $\mathcal{P}_-$ for $Q=M$ and $\mu=0.05\,M^2$. The bottom panel displays the corresponding difference image for  $Q=1.0115M$ and $\mu=0.05\,M^2$, representing an extreme EH BH solution. From Fig.~\ref{Difference_Image}, we conclude that there are subtle differences in the shadow and the gravitational lensing for the different polarizations of light propagating in the EH spacetime, mainly in the strong-field regime. In the weak-field limit, both polarizations lead to the same results, as discussed in Appendix~\ref{apxb}. Thus, the presence of the vacuum birefringence phenomenon in the context of NED gives rise to two different shadows and distinct gravitational lensing  for a single EH BH, depending on the observed polarization.  This result is important to the correct interpretation of the BH images obtained by the EHT Collaboration in the recent years~\cite{EHT2019, EHT2022}. It is important to notice that the polarization of light emitted around BHs has been recently measured also by the EHT collaboration~\cite{EHT2021}. However, a much higher resolution would be required to detect the difference between the images obtained for the two polarizations of the NED model studied here.

\subsection{Comparison with astrophysical data for Sgr A$^{\star}$}

To explore the potential relevance of the results presented here from the astrophysical point of view, we use the data of the instruments/teams Keck~\cite{TD2019} and the Very Large Telescope Interferometer (VLTI)~\cite{RA2020} for Sgr A$^{\star}$. According to them, the bounds in the shadow radius consistent with the most updated observations are~\cite{SV2023}
\begin{equation}
\label{1sigma}4.55 \lesssim r_{s}/M \lesssim 5.22,
\end{equation}
following $1\sigma$ constraints, and
\begin{equation}
\label{2sigma}4.21 \lesssim r_{s}/M \lesssim 5.56,
\end{equation}
for $2\sigma$ constraints. By drawing comparisons between the theoretical and observational results, we can impose constraints to the BH charge-to-mass ratio and to the EH parameter $\mu/M^{2}$.

\begin{figure}[!htbp]
\begin{centering}
    \includegraphics[width=\columnwidth]{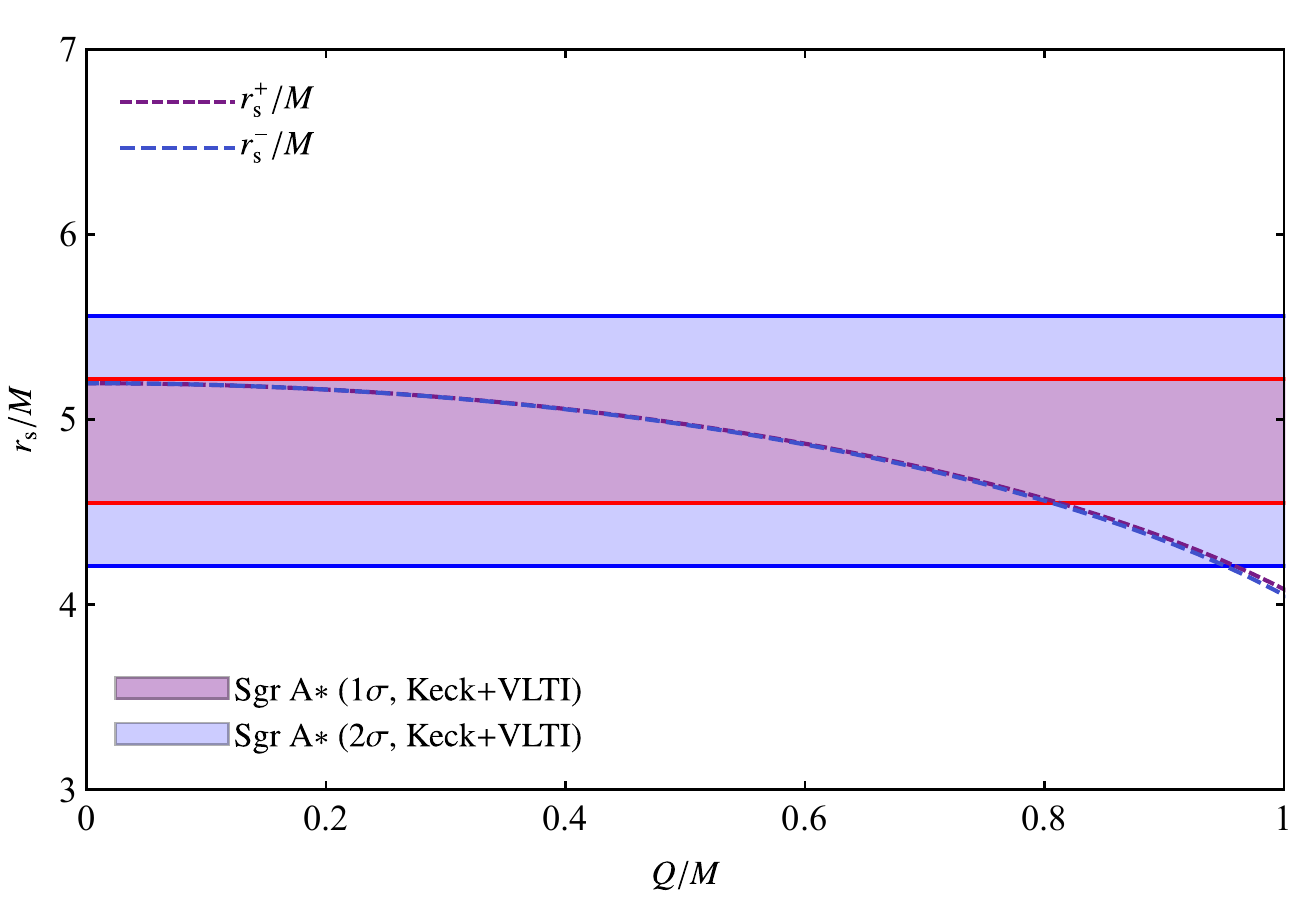}
    \caption{Shadow radius of the EH BH, considering the two polarizations of light, as functions of $Q/M$. The light blue and purple regions are consistent with the EHT horizon-scale image of Sgr A$^{\star}$ at $1\sigma$ and $2\sigma$, respectively. The white regions denote the excluded regions. The Keck~\cite{TD2019} and Very Large Telescope Interferometer (VLTI)~\cite{RA2020} are the instruments/teams that inferred, for example, the mass of Sgr A$^{\star}$. For more details, we recommend Ref.~\cite{SV2023}.}
    \label{sgt1}
\end{centering}
\end{figure}

We first compare the shadow radius computed according to the two polarizations of light with the observational data for Sgr A$^{\star}$, as shown in Fig.~\ref{sgt1}. We note that the astrophysical observations set the $1\sigma$ upper limits $Q \lesssim 0.8119M$ and $Q \lesssim 0.806M$ for $\mathcal{P}^{+}$ and $\mathcal{P}^{-}$, respectively. On the other hand, the $2\sigma$ constraints lead to the $2\sigma$ upper limits $Q \lesssim 0.9586M$ and $Q \lesssim 0.9503M$ for $\mathcal{P}^{+}$ and $\mathcal{P}^{-}$, respectively. Therefore, the observational data rules out the possibility of Sgr A$^{\star}$ being an extremal EH BH, as it occurs for the RN case, assuming $\mu = 0.05M^{2}$. In SI units, considering that $M = M_{\rm{Sgr A*}} = 4.297\times 10^{6}M_{\odot}$~\cite{RA2023}, the $2\sigma$ constraint on the EH BH charge leads to $Q \approx 6.99802\times 10^{26}C$. This charge value is substantially stronger than the limit $Q \lesssim 3\times 10^{8}C$, in units of Sgr A$^{\star}$ mass, obtained from the astrophysical considerations presented in Refs.~\cite{MZ2018,MZ2019}. These results are similar to those of RN~\cite{SV2023}.

\begin{figure}[!htbp]
\begin{centering}
    \includegraphics[width=\columnwidth]{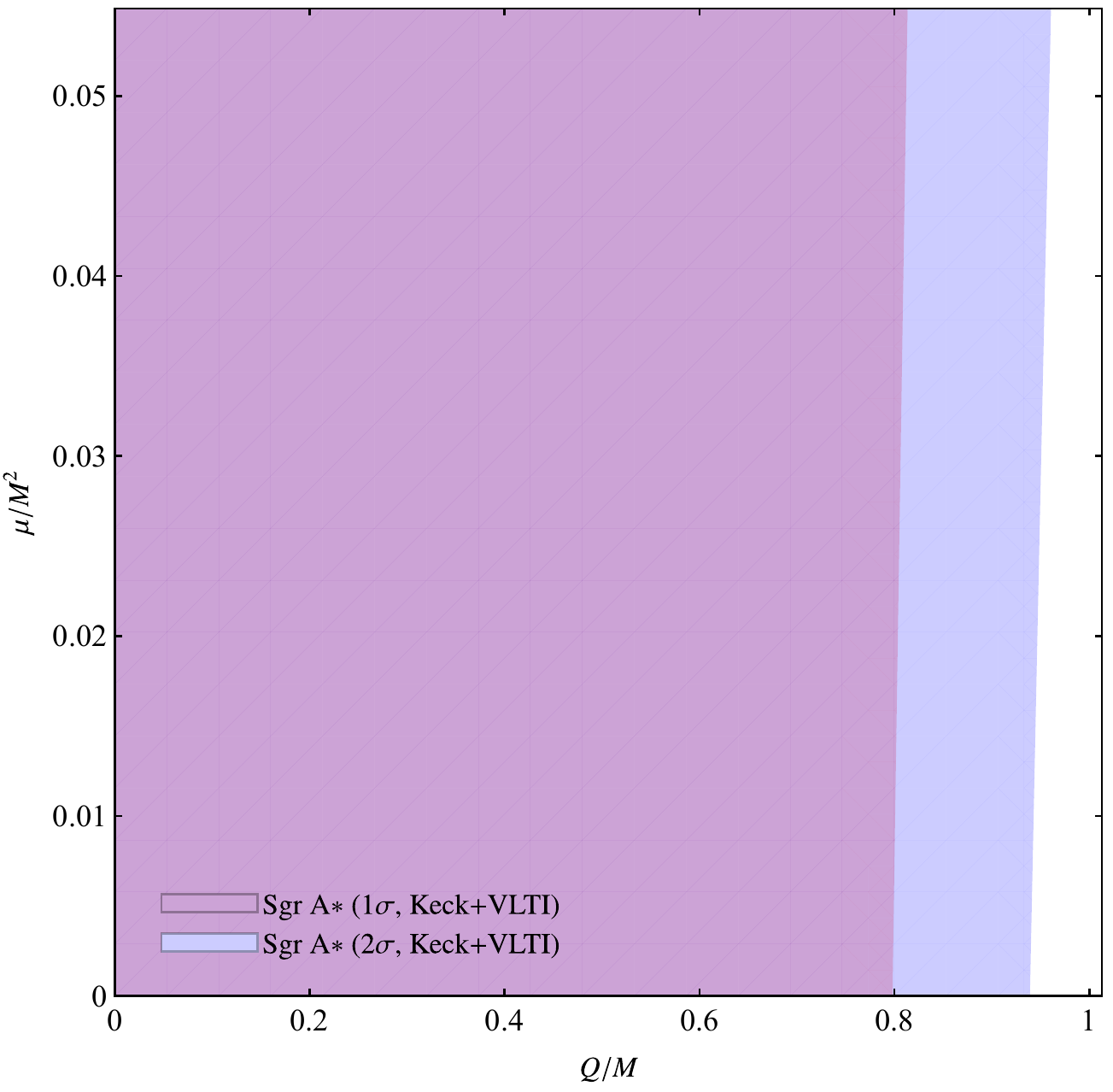}
    \caption{Allowed values of the EH parameter $\mu/M^{2}$ and BH charge-to-mass ratio $Q/M$, based on the observational results for Sgr A$^{\star}$. For simplicity, we consider the first light polarization, namely, $\mathcal{P}^{+}$.}
    \label{sgt2}
\end{centering}
\end{figure}

In Fig.~\ref{sgt2}, we constrain the values of $\mu/M^{2}$ and $Q/M$ based on Eqs.~\eqref{1sigma} and~\eqref{2sigma} for the polarization $\mathcal{P}^{+}$. The parameter space formed by $(\mu/M^{2},Q/M)$ shows that for $\mu/M^{2} = [0,40/729]$, we cannot have extremely charged EH BHs consistent with the astrophysical observations for the Sgr A$^{\star}$. The results for the polarization $\mathcal{P}^{-}$ are very similar to the results for the polarization $\mathcal{P}^{+}$, since the difference between the shadow radius of the two polarizations increases up to roughly $1\%$ (see, e.g., the bottom panel of Fig.~\ref{srr}).

It is also interesting to check whether the allowed values for $\mu$, considering the astrophysical observations for Sgr A$^{\star}$, are reasonable from the point of view of the EH electrodynamics \textit{per se}. As pointed out in Sec.~\ref{EH_Model_Sec}, the parameter $\mu$, in the context of the BH literature, is interpreted as a free parameter of the EH model that we vary to obtain BH solutions, although in the context of quantum electrodynamics it is fixed, as described by Eq.~\eqref{finestructure}. If we consider a typical value used for $\mu$ throughout this paper, e.g., $\mu = 0.05M^{2}$, we see that, in units of $M_{10k\odot}$, we have $\mu \approx  9.2 \times 10^{3}\left(M_{10k\odot}\right)^{2}$. This value is about one hundred thousand higher than the corresponding constant in quantum electrodynamics. It is necessary an extremely large value of $\mu$ to produce observable effects around a BH with mass comparable to Sgr A$^{\star}$.

We may reverse the reasoning and determine the range of BH masses for which the contributions become comparable to the quantum electrodynamics case given in Eq.~\eqref{finestructure}, assuming a fixed value $\mu = 0.05M^{2}$. In units of $M_{10k\odot}$, we obtain $M \approx 1.2 M_{10k\odot}$. This result indicates that the scale at which NED and strong gravity become comparable is astrophysical.

\section{Final Remarks}\label{sec:fr}

One of the main features of NED is that photons propagate along null geodesics of an effective geometry, and this geometry provides a valuable mechanism to reveal some imprints of nonlinear electromagnetic fields in the context of BH physics. In particular, it is of utmost importance to investigate the theoretical predictions of BHs sourced by NED for the gravitational waves and the BH shadows and to compare with the recent data obtained by several collaborations. In this work, we investigated the phenomenon of vacuum birefringence in the context of NED, focusing on the implications to the motion of photons around charged BHs. 

In NED models that depend on both  scalar invariants $F$ and $G$, photons can follow different paths, described by two distinct effective metrics, each one corresponding to a specific polarization. This vacuum birefringence effect gives rise to several interesting results for the motion of photons around BHs. We showed that the motion of photons in the background of the two effective metrics can be interpreted as nongeodesic curves subjected to four-force terms. This result extends the analysis performed in Ref.~\cite{MAP2024d}, where the four-force term was derived for a NED model depending solely on a single electromagnetic scalar invariant, specifically the invariant $F$. The four-force term emerges from the nonlinearities of the electromagnetic theory and behaves similarly to a Lorentz force, but acting on photons instead of charged particles. This interpretation, where photons follow nongeodesic curves under the influence of a four-force, is complementary to the effective metric approach.

We also investigated optical effects around these BHs. Due to the two different effective geometries, photons experience different gravitational lensing, depending on their polarizations. LRs result from strong gravitational lensing around BHs, where photons describe closed circular orbits. Due to the presence of two effective metrics, a static BH sourced by NED can exhibit two distinct unstable LRs, one for each polarization, and no stable LRs. This configuration where a BH casts two unstable LRs and no stable LR circumvents the theorem proved recently for the topological charge number associated with LRs~\cite{PV:2020}. However, with a few straightforward modifications, the theorem can be extended to incorporate the two distinct effective metrics governing photon motion, which constitutes a possible extension of this work.

As an application of the general results discussed in this work, we studied the LRs, shadows and gravitational lensing for the EH BH which is an example of a BH sourced by NED that depends on the two scalar invariants. We obtained that a single EH BH solution can cast two distinct unstable LRs, and although these unstable LRs do not coincide, the values of their radial coordinates are similar. LRs play an important role in the investigation of the BH image, as their associated impact parameter defines the boundary of the shadow. The existence of two unstable LRs gives rise to two distinct shadows of a single EH BH, depending on the observed polarization. In order to investigate the shadow and the gravitational lensing of the EH BH solution, we applied the backwards ray-tracing technique for both effective geometries. The backwards ray-tracing simulations showed that the shadow and the gravitational lensing are very similar for the two polarizations, but subtle differences can be highlighted by computing a grayscale difference image, obtained through a pixelwise comparison of the two images.

We notice that, although our results show that the polarization-induced splitting in the photon sphere and shadow radii is relatively small, the corresponding splitting of the shadow radius constitutes, at least in principle, a genuine observable signature of NED effects. At present, the predicted deviations are likely below the sensitivity of current observational facilities, such as the EHT. However, the purpose of our analysis is primarily theoretical: to quantify how polarization-dependent propagation modifies the optical appearance of NED-based BHs. In this sense, the results provide a benchmark for future high-precision observations and for assessing whether next-generation instruments could eventually probe such effects.

We also aimed to explore the potential astrophysical relevance of the results presented here by drawing comparisons with the EHT observation data for Sgr A$^{\star}$. Considering up to $2\sigma$ constraints, we have shown that the EHT observations rule out the possibility of Sgr A$^{\star}$ being an extremal EH BH, as it occurs for the RN case, assuming $\mu/M^{2} = [0,40/729]$. Moreover, we obtained that the $2\sigma$ constraints on the BH charge leads to $Q \approx 6.99802\times 10^{26}C$, in units of Sgr A$^{\star}$ mass. We have also found that, considering only the $1\sigma$ constraint, not even moderately-charged EH BHs are allowed in view of EHT observations. We have restricted our results to BHs only.

The theoretical predictions for the shadow and gravitational lensing presented in this work are particularly relevant in light of recent results from the EHT collaboration. Here, we have explored only the $1\sigma$ and $2\sigma$ constraints related to the EHT observational data associated with Sgr A$^{\star}$ and considered spherically symmetric BHs. As potential extensions of this study, we can explore rotating BHs sourced by NED and examine the vacuum birefringence phenomenon in this context.

\begin{acknowledgments}

We acknowledge Funda\c{c}\~ao Amaz\^onia de Amparo a Estudos e Pesquisas (FAPESPA), Conselho Nacional de Desenvolvimento Cient\'ifico e Tecnol\'ogico (CNPq) and Coordena\c{c}\~ao de Aperfei\c{c}oamento de Pessoal de N\'ivel Superior (CAPES) -- Finance Code 001, from Brazil, for partial financial support. This work is supported by CIDMA under the Portuguese Foundation for Science and Technology (FCT, \url{https://ror.org/00snfqn58}) Multi-Annual Financing Program for R\&D Units, Grants No. UID/4106/2025,  No. UID/PRR/4106/2025”, and No. 2022.04560.PTDC (with the DOI https://doi.org/10.54499/2022.04560.PTDC), 2024.05617.CERN (https://doi.org/10.54499/2024.05617.CERN). This work has further been supported by the European Horizon Europe staff exchange (SE) programme HORIZON-MSCA-2021-SE-01 Grant No. NewFunFiCO-101086251. M. P. thanks Sérgio Xavier for useful discussions on the PyHole Python package. L. C. and H. C. D. L. would like to thank the University of Aveiro, in Portugal, for the kind hospitality during the completion of this work. M. A. A. de Paula is supported by CNPQ/PDJ 150589/2025-5.

\end{acknowledgments}

\section*{Data availability}

There are no publicly available research data or software supporting this manuscript. Requests for further information or data should be sent to the authors.

\appendix

\section{P Framework}\label{apxA}

In this appendix, we show that the results obtained in Sec.~\ref{sec:eg4f} successfully reproduce those presented in Appendix B of Ref.~\cite{MP2023}, considering the appropriate limits. In NED, it is usual to introduce an auxiliary antisymmetric tensor to obtain exact electrically charged solutions of the ENED field equations [cf. Eq.~\eqref{E-NED_F}] in the presence of nonlinear electromagnetic sources~\cite{HGP1987}. The so-called $P$ \textit{framework} is introduced through a Legendre transformation~\cite{HGP1987}, namely
\begin{equation}
\label{LT_H}\mathcal{H}(P) \equiv 2F\mathcal{L}_{F} - \mathcal{L}(F),
\end{equation}
where $\mathcal{H}(P)$ is the structural function. The auxiliary antisymmetric tensor $P_{\mu\nu}$ and the scalar $P$ are given by
\begin{equation}
\label{AT}P_{\mu\nu} = \mathcal{L}_{F}F_{\mu\nu} \quad \text{and} \quad P = P_{\mu\nu}P^{\mu\nu},
\end{equation}
respectively. With Eqs.~\eqref{LT_H} and~\eqref{AT}, we can show that
\begin{equation}
\label{FPDUALITY}P = (\mathcal{L}_{F})^{2}F, \ \ \mathcal{H}_{P}\mathcal{L}_{F} = 1\ \ \text{and} \ \ F_{\mu\nu} = \mathcal{H}_{P}P_{\mu\nu},
\end{equation}
where $\mathcal{H}_{P} \equiv \partial\mathcal{H}/\partial P$, and Eq.~\eqref{LT_H} can be rewritten as
\begin{equation}
\label{LT_L}\mathcal{L}(F) = 2P\mathcal{H}_{P} - \mathcal{H}(P).
\end{equation}
This framework fulfills a correspondence with Maxwell's theory in the weak field limit if $\mathcal{H}(P) \rightarrow P$ and $\mathcal{H}_{P} \rightarrow 1$, for small $P$. Moreover, the connection between the $F$ and $P$ representations of NED is called FP duality. Simply put, in the spherically symmetric case, any electric (magnetic) solution obtained in the $F$ \textit{framework} has a magnetic (electric) counterpart in the $P$ \textit{framework} and vice versa~\cite{B2001,B2018}. Thus, distinct NED models can be connected via the same line element. The FP duality turns into the standard electric-magnetic duality only in the linear electrodynamics case.

By using Eqs.~\eqref{FPDUALITY}, we can find that
\begin{equation}
\label{LffHpp}\mathcal{L}_{FF} = - \dfrac{\mathcal{H}_{PP}}{\mathcal{H}_{P}^{2}F_{P}}.
\end{equation}
With the help of the Eqs.~\eqref{AT},~\eqref{LT_L}, and~\eqref{LffHpp}, we can show that Eq.~\eqref{4force} reduces to Eq.~(B6) of Ref.~\cite{MP2023}. Therefore, we may interpret the results presented by the authors in the Appendix B of Ref.~\cite{MP2023} as a particular case of the results obtained here.

\section{Weak Deflection Angle Via Gauss-Bonnet Theorem}\label{apxb}

In this Appendix, we derive an expression for the weak deflection angle of the magnetically charged EH BH by means of the Gauss-Bonnet theorem~\cite{GW2008}, considering the standard and effective geometries. This approach has been successfully used in several spherically symmetric NED-based BH setups (see, e.g., Refs.~\cite{O2019,JAO2019,JHO2020,FZL2021,OA2022} and references therein). The method is also used in more general scenarios~\cite{AI2016,OIA2017,JV2018}, and its relations with other methods have been discussed in depth in other works~\cite{HCL2024} (see also references therein).

The line element of the optical metric related to the motion of massless particles can be obtained by setting $ds^{2} = 0$ and $\theta = \pi/2$ in the line element~\eqref{LE}. Thus, we get
\begin{equation}
\label{LE_OM}dt^{2} = g_{\mu\nu}^{\rm{opt}}dx^{\mu}dx^{\nu} = \dfrac{dr^{2}}{f(r)^{2}}+\dfrac{r^{2}d\varphi^{2}}{f(r)}.
\end{equation}
By using the tortoise coordinate, given by $dr = f(r)dr_{\star}$, and defining $v(r_{\star}) = r/\sqrt{f(r)}$, we can rewrite Eq.~\eqref{LE_OM} as
\begin{equation}
\label{LE_OM2}dt^{2} = h_{\mu\nu}^{\rm{opt}}dx^{\mu}dx^{\nu} = dr_{\star}^{2}+v(r_{\star})^{2}d\varphi^{2}.
\end{equation}
The Gaussian optical curvature $\mathcal{K}$ is given by
\begin{equation}
\label{goc}\mathcal{K} = \dfrac{R}{2} = -\dfrac{1}{v(r_{\star})}\dfrac{d^{2}v(r_{\star})}{dr_{\star}^{2}},
\end{equation}
where $R$ is the Ricci scalar associated with the geometry~\eqref{LE_OM2}.

By using the Gauss-Bonnet theorem, we can express the deflection angle $\Theta(b)$ as given by~\cite{GW2008}
\begin{equation}
\label{DA_GBT}\Theta(b) = - \int_{0}^{\pi + \alpha}\int_{u(\varphi)}^{\infty}\mathcal{K}dS,
\end{equation}
where
\begin{equation}
dS = \sqrt{|h^{\rm{opt}}|}dr_{\star}d\varphi,
\end{equation}
with $h^{\rm{opt}}$ being the determinant of the metric tensor $h_{\mu\nu}^{\rm{opt}}$. The parameters $\alpha$ and $u(\varphi)$ are the corrections in the observer's position and the trajectory of the particle, obtained by solving the orbit equation~\eqref{RE_EG}. For our purposes, we can set the correction as $\alpha = 4M/b$, and suppose that $u(\varphi)$ is given by
\begin{equation}
\label{pt}u(\varphi) = \dfrac{1}{b}\left[u_{0}(\varphi)+\dfrac{M}{b}u_{1}(\varphi) + \dfrac{M^{2}}{b^{2}}u_{2}(\varphi)\right],
\end{equation}
where the coefficients $u_{k}(\varphi)$ can be obtained via an iterative method~\cite{AK2012}.

Applying the same approach for the effective metrics~\eqref{LE_EF}, the corresponding function $v(r_{\star})$ leads to
\begin{equation}
\label{funcv}v_{\pm}(r_{\star})^{2} = \dfrac{r^{2}G^{\pm}_{1}(r)}{G^{\pm}_{2}(r) f(r)}.
\end{equation}
Hence, the equations we need to use to compute the deflection angles in the weak-field limit, considering the effective metrics, are the same, provided that we replace $v(r_{\star})$ with $v_{\pm}(r_{\star})$.

An explicit calculation shows that the weak deflection angle of the standard and effective metrics coincide up to the third order in $1/b^{3}$, resulting in
\begin{align}
\nonumber \Theta (b) = \  & \dfrac{4M}{b} + \dfrac{3\pi \left(5M^{2}-Q^{2}\right)}{4b^{2}} + \\
&\dfrac{16M\left(8M^{2}-3Q^{2}\right)}{3b^{3}} + \mathcal{O}\left[\dfrac{1}{b^{4}}\right],
\end{align}
which coincides with the RN result~\cite{KP2005}, and for $Q = 0$ we obtain the Schwarzschild one. The similarity with the RN case can be understood by noting that the charge contributions associated with the EH BH are proportional to $1/r^{6}$ in the metric function and to $1/r^{4}$ in the magnetic factors. Therefore, in order to search for perturbations arising from the EH source in the weak deflection angle, we need to consider more correction orders, which is beyond the scope of this work.

\end{document}